\documentclass{article}%
\usepackage{amsmath}
\usepackage{amsfonts}
\usepackage{amssymb}
\usepackage{graphicx}%
\begin{document}

\title{Quantitative Simulation of the Superconducting Proximity Effect }
\date{\today}
\author{Gerd Bergmann\\Physics Department, Univ.South.California\\Los Angeles, CA 90089-0484, USA}
\maketitle

\begin{abstract}
A numerical method is developed to calculate the transition temperature of
double or multi-layers consisting of films of super- and normal conductors.
The approach is based on a dynamic interpretation of Gorkov's linear gap
equation and is very flexible. The mean free path $l$ of the different metals,
transmission through the interface, ratio of specular reflection to diffusive
scattering at the surfaces, and fraction of diffusive scattering at the
interface can be included. Furthermore it is possible to vary the mean free
path and the BCS interaction $NV$ in the vicinity of the interface. The
numerical results show that the normalized initial slope of an SN double layer
is independent of almost all film parameters except the ratio of the density
of states, $\left(  d_{s}/T_{s}\right)  \left\vert dT_{s}/dd_{n}\right\vert
=\Gamma_{sn}\left(  N_{n}/N_{s}\right)  $. There are only very few
experimental investigations of this initial slope and they consist of Pb/Nn
double layers (Nn stands for a normal metal). Surprisingly the coefficient
$\Gamma_{sn}$ in these experiments is of the order or less than 2 while the
(weak coupling) theory predicts a value of about 4.5. This discrepancy has not
been recognized in the past. The autor suggests that it is due to strong
coupling behavior of Pb in the double layers. The strong coupling gap equation
is evaluated in the thin film limit and yields the value of 1.6 for
$\Gamma_{sn}$. This agrees much better with the few experimental results that
are available.


\newpage

\end{abstract}

\section{Introduction}

The transition temperature of a thin superconducting film in contact with a
normal metal is reduced. This is known as the superconducting proximity effect
(SPE). The double layer SN or a multi layer (SN)$_{n}$ can consist (i) of a
superconductor S and normal conductor N or (ii) of two superconductors with
different transition temperatures (the one with the lower $T_{c}$ is generally
denoted as N). Its systematic experimental investigation started in 1960's by
the Hilsch group in Goettingen \cite{H31}, \cite{H26} and stimulated a number
of further experimental investigations \cite{D36}, \cite{H27}, \cite{B133},
\cite{M45}. For the dirty case (mean free path of the conduction electrons is
much smaller than the coherence length) Werthamer \cite{W32} derived a set of
implicit equations for the transition temperature of double layers consisting
of two superconductors. After some modification according to de Gennes'
boundary condition \cite{D12} between the superconductors, the Wertheimer
theory described the experimental results for double layers of two
superconductors quite well (see for example \cite{B133}, \cite{D12},
\cite{D32}). The Wertheimer theory is restricted to short mean free path,
(using the diffusion approximation) and uses what is now called the single
mode approximation (the gap function is approaximated by a $\cos\left(
k_{s}z\right)  $-dependence). Theoretical results for the clean case where the
mean free path $l$ is larger than the BCS coherence length $\xi_{BCS}$ are
more difficult and the case where $l,\xi_{BCS}$ and the film thicknesses are
of the same order of magnitude are much more challenging.

In recent years the superconducting proximity effect has experienced a renewed
interest. A large number of papers studied the SPE theoretically \cite{A63},
\cite{B161}, \cite{A62}, \cite{A57}, \cite{Z7}, \cite{Z8}, \cite{V11},
\cite{S39}, \cite{S50}, \cite{S40}, \cite{S51}, \cite{T16}, \cite{N13},
\cite{Z9} and experimentally \cite{N9}, \cite{V14}, \cite{B144}, \cite{P31},
\cite{S39}, \cite{M62}, \cite{S51}, \cite{D40}, \cite{B134}, \cite{B135},
\cite{V16} particularly during the last 10 years. The studies have been
extended to double layers of a superconductor and a ferromagnet (SF)
\cite{G41}, \cite{B160}, \cite{K54}.

Recently our group revisited the superconducting proximity effect using it as
an experimental tool \cite{B135}, \cite{B134}. One interesting information the
SPE provides is the transparency of the interface between the two metal films
for the conduction electrons. The reduction of $T_{c}$ in the superconducting
component of the SN double layer depends on the rate at which electrons can
cross the interface between S and N. This interface transparency is of
interest in a number of other disciplines and applications in solid state physics.

When our group tried to compare the experimental results for the transition
temperature with theoretical predictions we found that only a few recent
theoretical investigations calculated the transition temperature of SN double
layers \cite{A57}, \cite{B161}, \cite{F34}. These papers considered the
extreme cases, either the clean limit for infinitely large mean free path
\cite{A57} \ or the dirty limit \cite{F34} where the mean free path is much
shorter than the BCS coherence length. Reference \cite{F34} considered
superconductor-ferromagnet double layers in the "dirty limit". It includes the
case of an SN double layer by setting the exchange energy in the ferromagnet
equal to zero. A multi-mode expansion of the order parameter is used in the
superconductor. This yields a complex set of equations which contain the
transition temperature implicitely. Their single mode approximation is similar
to Wertheimers result.

Since our experiments used films with short and large mean free paths the
author prefered to develop a numerical procedure which is capable of
calculating the transition temperature of arbitrary sequences of
superconductors and normal conductors in a wide range of the mean free path.
This calculation uses a simple interpretation of the gap equation which was
stimulated by deGennes work \cite{D12}. Below I will sketch the (simple)
numerical procedure. In chapter II the theoretical background is reviewed and
the numerical procedure discussed in detail. In chapter III some of the
numerical results are presented. In the discussion of chapter IV I will point
out a discrepancy between all experiments I am aware of which study the change
of $T_{c}$ of a superconducting film when covered with a thin normal
conducting film, i.e. the normalized intial slope $\dfrac{d_{s}}{T_{s}}%
\dfrac{dT_{c}}{dd_{n}}$, where $d_{s}$ and $T_{s}$ are the thickness and
transition tmperature of the superconductor and $\dfrac{dT_{c}}{dd_{n}}$ is
the initial slope of the $T_{c}$-reduction for zero thickness $d_{n}$ of the
normal conductor.\newpage

\section{Theoretical Background}

\subsection{The linear gap equation}

The superconducting phase transition in zero magnetic field is generally of
second order. Therefore, close to transition temperature $T_{c}$ of the double
layer, the gap function $\Delta\left(  \mathbf{r}\right)  $ , which is the
order parameter of the phase transition, is small and only terms linear in the
gap function contribute. This linear gap equation, first formulated by Gorkov
\cite{G38} was rewritten by deGennes \cite{D12} as%

\begin{align}
\Delta\left(  \mathbf{r}\right)   & =V\left(  \mathbf{r}\right)  \int
d^{3}\mathbf{r}^{\prime}\sum_{\left\vert \omega_{n}\right\vert <\Omega_{D}%
}^{n_{c}}H_{\omega_{n}}\left(  \mathbf{r,r}^{\prime}\right)  \Delta\left(
\mathbf{r}^{\prime}\right) \label{gap0}\\
H_{\omega_{n}}\left(  \mathbf{r,r}^{\prime}\right)   & =k_{B}TG_{\omega_{n}%
}\left(  \mathbf{r,r}^{\prime}\right)  G_{\omega_{n}}^{\ast}\left(
\mathbf{r,r}^{\prime}\right) \nonumber
\end{align}

Here $\Delta\left(  \mathbf{r}\right)  $ is the gap function at the position
$\mathbf{r}$, $\omega_{n}=\left(  2n+1\right)  \pi k_{B}T/\hbar$ are the
Matsubara frequencies, $V\left(  \mathbf{r}\right)  $ is the effective
electron-electron interaction at the position $\mathbf{r}$. The sum is limited
to the range $\left\vert \omega_{n}\right\vert <\Omega_{D}$ where $\Omega_{D}$
is the Debye temperature. This corresponds to a sum over $n$ from $-n_{c}$ to
$+n_{c}$ where $n_{c}=\Theta_{D}/\left(  2\pi T\right)  =\Omega_{D}\tau_{T},$
where $\Theta_{D}$ and $\ \Omega_{D}$ are the Debye temperature and frequency
and $\tau_{T}=\hbar/\left(  2\pi k_{B}T\right)  $. The function $H_{\omega
_{n}}\left(  \mathbf{r,r}^{\prime}\right)  $ is the product of two Green
functions $G_{\omega_{n}}\left(  \mathbf{r,r}^{\prime}\right)  $ and
$G_{\omega_{n}}^{\ast}\left(  \mathbf{r,r}^{\prime}\right)  $ which represent
a Cooperon. Since the Green function $G_{\omega_{n}}\left(  \mathbf{r,r}%
^{\prime}\right)  $ represents the amplitude of an electron traveling (at
finite temperarture) from $\mathbf{r}^{\prime}$ to $\mathbf{r}$ the product
$G_{\omega_{n}}\left(  \mathbf{r,r}^{\prime}\right)  G_{\omega_{n}}^{\ast
}\left(  \mathbf{r,r}^{\prime}\right)  $ describes the amplitude of a Cooperon
traveling from $\mathbf{r}^{\prime}$ to $\mathbf{r}$. Since the two
single-particle Green functions are conjugate complex to each other, the
product of their amplitudes is proportional to the probability of a single
electron to travel from $\mathbf{r}^{\prime}$ to $\mathbf{r}$. If one
interprets in equation $\left(  \ref{gap0}\right)  $ $G_{\omega_{n}}\left(
\mathbf{r,r}^{\prime}\right)  G_{\omega_{n}}^{\ast}\left(  \mathbf{r,r}%
^{\prime}\right)  $ as the propagation of single electrons then one has an
equivalent problem and its solution is also the solution of the gap equation.
In the following the solution of the equivalent problem will be persued.

From the properties of the Green functions $G_{\omega_{n}}\left(
\mathbf{r,r}^{\prime}\right)  $ (see appendix 6.1) it follows that
$H_{\omega_{n}}\left(  \mathbf{r,r}^{\prime}\right)  $ is the electron density
if one injects continuously electrons with a rate $N/\tau_{T}$ at the point
$\mathbf{r}^{\prime},$ while their density decays along the path as
$\exp\left(  -2\left\vert \omega_{n}\right\vert s/v_{F}\right)  $ where $s$ is
the distance traveled (not the distance from $\mathbf{r}^{\prime}$) and $N$ is
the BCS density of states.

The right side of equation $\left(  \ref{gap0}\right)  $ $d^{3}\mathbf{r}%
^{\prime}H_{\omega_{n}}\left(  \mathbf{r,r}^{\prime}\right)  \Delta\left(
\mathbf{r}^{\prime}\right)  $ (excluding $\sum_{n=-n_{c}}^{n_{c}}$) yields the
density of electrons at the position $\mathbf{r}$ when one injects constantly
$N\Delta\left(  \mathbf{r}^{\prime}\right)  d^{3}\mathbf{r}^{\prime}%
dt^{\prime}/\tau_{T}$ electrons in the incremental volume $d^{3}%
\mathbf{r}^{\prime}$ at the position $\mathbf{r}^{\prime}$ per time interval
$dt^{\prime},$ which decay during their propagation with the decay rate of
$2\left\vert \omega_{n}\right\vert $ ($\tau_{T}=\hbar/\left(  2\pi
k_{B}T\right)  $). ($N\Delta\left(  \mathbf{r}^{\prime}\right)  d^{3}%
\mathbf{r}^{\prime}$ represents a (dimensionless) number of electrons and the
rate of injected electrons per volume is $N\Delta\left(  \mathbf{r}^{\prime
}\right)  /\tau_{T}$). These electrons propagate with their Fermi velocity
from $\mathbf{r}^{\prime}$ to $\mathbf{r}$, either directly or diffusively.
Their density decays along the path as $\exp\left(  -2\left\vert \omega
_{n}\right\vert t_{\Delta}^{\prime}\right)  $ where $t_{\Delta}^{\prime}$ is
the time since the departure from $\mathbf{r}^{\prime}$. At the position
$\mathbf{r}$ the surviving density of all arriving electrons is integrated
over $\int d^{3}\mathbf{r}^{\prime}\int_{-\infty}^{0}dt^{\prime}$. When summed
over $\omega_{n}$ and multiplied with the attractive electron interaction
$V\left(  \mathbf{r}\right)  $ one has to recover the original $\Delta\left(
\mathbf{r}\right)  $.

For further treatment we define the propagation density $\rho\left(
v_{F};\mathbf{r,}0;\mathbf{r}^{\prime},t^{\prime}<0\right)  $. If an electron
with Fermi velocity $v_{F}$ is introduced at the time $t^{\prime}<0$ at the
position $\mathbf{r}^{\prime}$ then $\rho\left(  v_{F};\mathbf{r,}%
0;\mathbf{r}^{\prime},t^{\prime}\right)  $ describes the probability to find
the electron at the time $0$ at the position $\mathbf{r}$. With this
definition we can express $H_{\omega_{n}}\left(  \mathbf{r},\mathbf{r}%
^{\prime}\right)  $%
\[
H_{\omega_{n}}\left(  \mathbf{r},\mathbf{r}^{\prime}\right)  =N\left(
\mathbf{r}^{\prime}\right)  \int_{-\infty}^{0}\rho\left(  v_{F};\mathbf{r,}%
0;\mathbf{r}^{\prime},t^{\prime}\right)  \exp\left(  -2\left\vert \omega
_{n}\right\vert \left\vert t^{\prime}\right\vert \right)  \dfrac{dt^{\prime}%
}{\tau_{T}}%
\]
where $1/\tau_{T}=2\pi k_{B}T/\hbar$.

The sum over $\omega_{n}$ in $\left(  \ref{gap0}\right)  $ applies only to the
exponential decay functions $\exp\left(  -2\left\vert \omega_{n}\right\vert
\left\vert t^{\prime}\right\vert \right)  $ and yields the time function
\ $\eta_{T}\left(  t^{\prime}\right)  $
\begin{equation}
\eta_{T}\left(  t^{\prime}\right)  =\sum_{\left\vert \omega_{n}\right\vert
<\Omega_{D}}\exp\left(  -\left\vert \omega_{n}\right\vert \left\vert
t^{\prime}\right\vert \right)  =\dfrac{1-\exp\left(  -2\left(  \Omega_{D}%
\tau_{T}+1\right)  \left\vert t^{\prime}\right\vert /\tau_{T}\right)  }%
{\sinh\left(  \left\vert t^{\prime}\right\vert /\tau_{T}\right)
}\label{dcayf}%
\end{equation}
($\Omega_{D}$=Debye frequency). Then one can express the gap equation as
\begin{equation}
\Delta\left(  \mathbf{r}\right)  =V\left(  \mathbf{r}\right)  \int
d^{3}\mathbf{r}^{\prime}N\left(  \mathbf{r}^{\prime}\right)  \int_{-\infty
}^{0}\dfrac{dt^{\prime}}{\tau_{T}}\eta_{T}\left(  t^{\prime}\right)
\rho\left(  v_{F};\mathbf{r,}0\mathbf{;r}^{\prime},t^{\prime}\right)
\Delta\left(  \mathbf{r}^{\prime}\right) \label{gap2a}%
\end{equation}

It is obvious that the superconducting properties of the system occur only in
the effective interaction $V\left(  \mathbf{r}^{\prime}\right)  $ and the
decay function $\eta_{T}\left(  t^{\prime}\right)  $. Of course,
$\Delta\left(  \mathbf{r}\right)  $ is the superconducting pair amplitude but
in equation $\left(  \ref{gap2a}\right)  $ it is just the eigen vector of the
integral kernel. The self-consistancy condition requires that this eigen value
is equal to one.

This interpretation of the gap equation yields a natural extension to a time
dependent pair amplitude or gap function. One obtains%
\begin{equation}
\Delta\left(  \mathbf{r},t\right)  =V\left(  \mathbf{r}\right)  \int
d^{3}\mathbf{r}^{\prime}N\left(  \mathbf{r}^{\prime}\right)  \int_{-\infty
}^{t}\dfrac{dt^{\prime}}{\tau_{T}}\eta_{T}\left(  t^{\prime}\right)
\rho\left(  v_{F};\mathbf{r,}0\mathbf{;r}^{\prime},t^{\prime}\right)
\Delta\left(  \mathbf{r}^{\prime},t^{\prime}\right) \label{gap2c}%
\end{equation}
From this equation one can derive a time dependent Ginsburg-Landau equation
\cite{B22}.

For a homogeneous superconductor one has a constant energy gap. In this case
one can perform the integral over $d^{3}\mathbf{r}^{\prime},$ using $\int
d^{3}\mathbf{r}\rho\left(  v_{F};\mathbf{r,}0;\mathbf{r}^{\prime},t^{\prime
}\right)  =1$ and dividing by $\Delta$%
\begin{equation}
1=\left(  VN\right)  _{s}\int_{-\infty}^{0}\dfrac{dt^{\prime}}{\tau_{T}}%
\eta_{T}\left(  t^{\prime}\right) \label{constge}%
\end{equation}
which yields%
\[
\frac{1}{NV}=\sum_{n=0}^{n_{c}}\frac{1}{n+\frac{1}{2}}%
\]

The condition (\ref{constge}) is used to determine the BCS coupling strength
$\left(  NV\right)  _{s}.$It has the advantage that it is not restricted to
integer values of $n_{c}=\Omega_{D}\tau_{T}$.

\subsection{The gap equation for double and multi-layers}

Now we can apply the gap equation $\left(  \ref{gap2a}\right)  $ to the
proximity effect. The direction $z$ is chosen perpendicular to the multi-layer
and the films are treated as homogeneous in the x-y plane. If there is no
magnetic field then the gap depends only on the z direction. Therefore one can
perform the integration over $\int dx^{\prime}dy^{\prime}\rho\left(
v_{F};\mathbf{r,}0\mathbf{;r}^{\prime},t^{\prime}\right)  =\overline{\rho
}\left(  z\mathbf{,}0\mathbf{;}z^{\prime},t^{\prime}\right)  $.

Now the function $\overline{\rho}\left(  z\mathbf{,}0\mathbf{;}z^{\prime
},t^{\prime}\right)  $ describes the density at the time $t=0$ and the
position $z$ integrated over the x and y directions. For the numerical
procedure it is more convenient to shift the time integration from the range
$\left(  -\infty,0\right)  $ to the range $\left(  0,\infty\right)  $.%

\[
\Delta\left(  z\right)  =V\left(  z\right)  \int dz^{\prime}N\left(
z^{\prime}\right)  \int_{0}^{\infty}\dfrac{dt}{\tau_{T}}\eta_{T}\left(
t\right)  \overline{\rho}\left(  z\mathbf{,}t\mathbf{;}z^{\prime},0\right)
\Delta\left(  z^{\prime}\right)
\]

The multi-layer will be divided into small sheets parallel to the film
surfaces. The layers are indexed by $\nu$ and possess a thickness
$\lambda_{\nu}$.
\[%
{\parbox[b]{3.6953in}{\begin{center}
\includegraphics[
height=2.5745in,
width=3.6953in
]%
{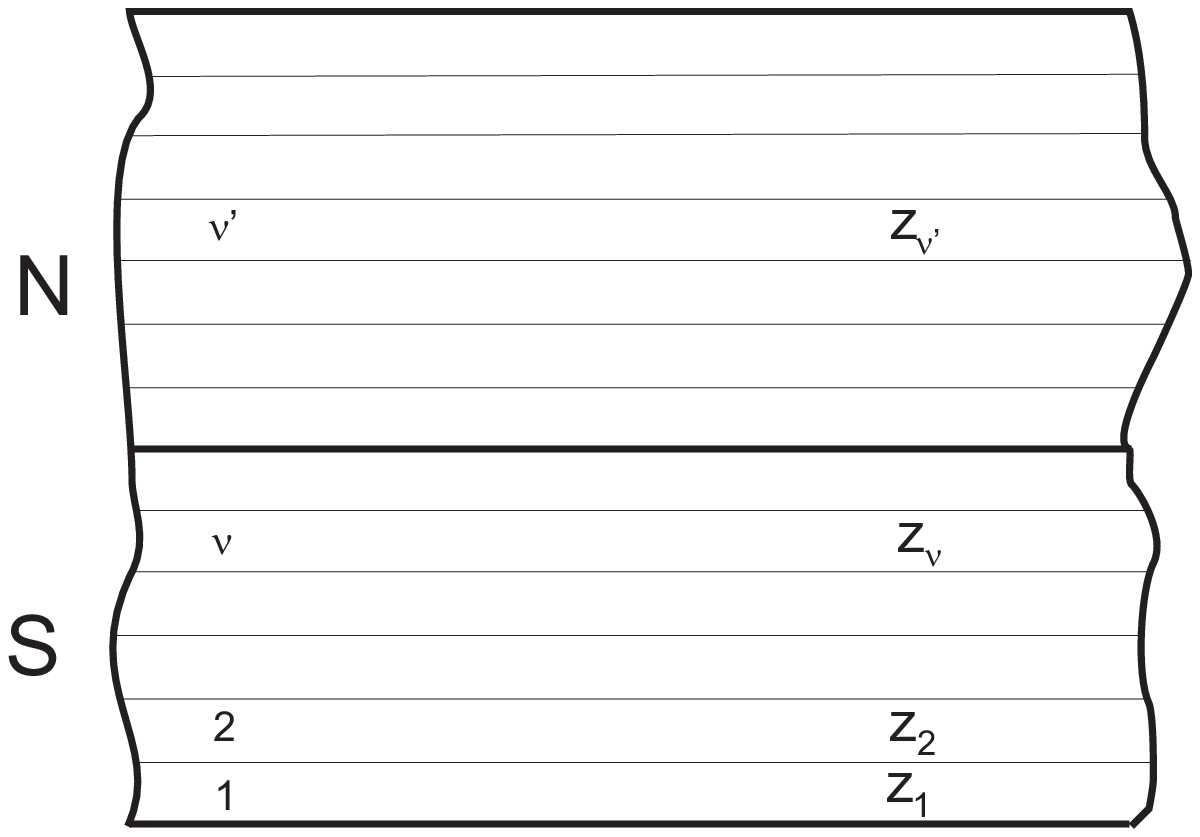}%
\\
Fig.1: A double layer of a superconductor S and a normal conductor N. The two
films are split in thin paralllel layers $\nu$ with the z position $z_\nu$
\end{center}}}%
\]

In the present paper we determine the gap-function $\Delta\left(  z\right)  $
at the transition temperature of an SN (superconductor/normal metal) double
layer. We proceed with the following steps which are demonstrated by Fig.2.%
\[%
{\parbox[b]{3.1042in}{\begin{center}
\includegraphics[
height=2.5745in,
width=3.1042in
]%
{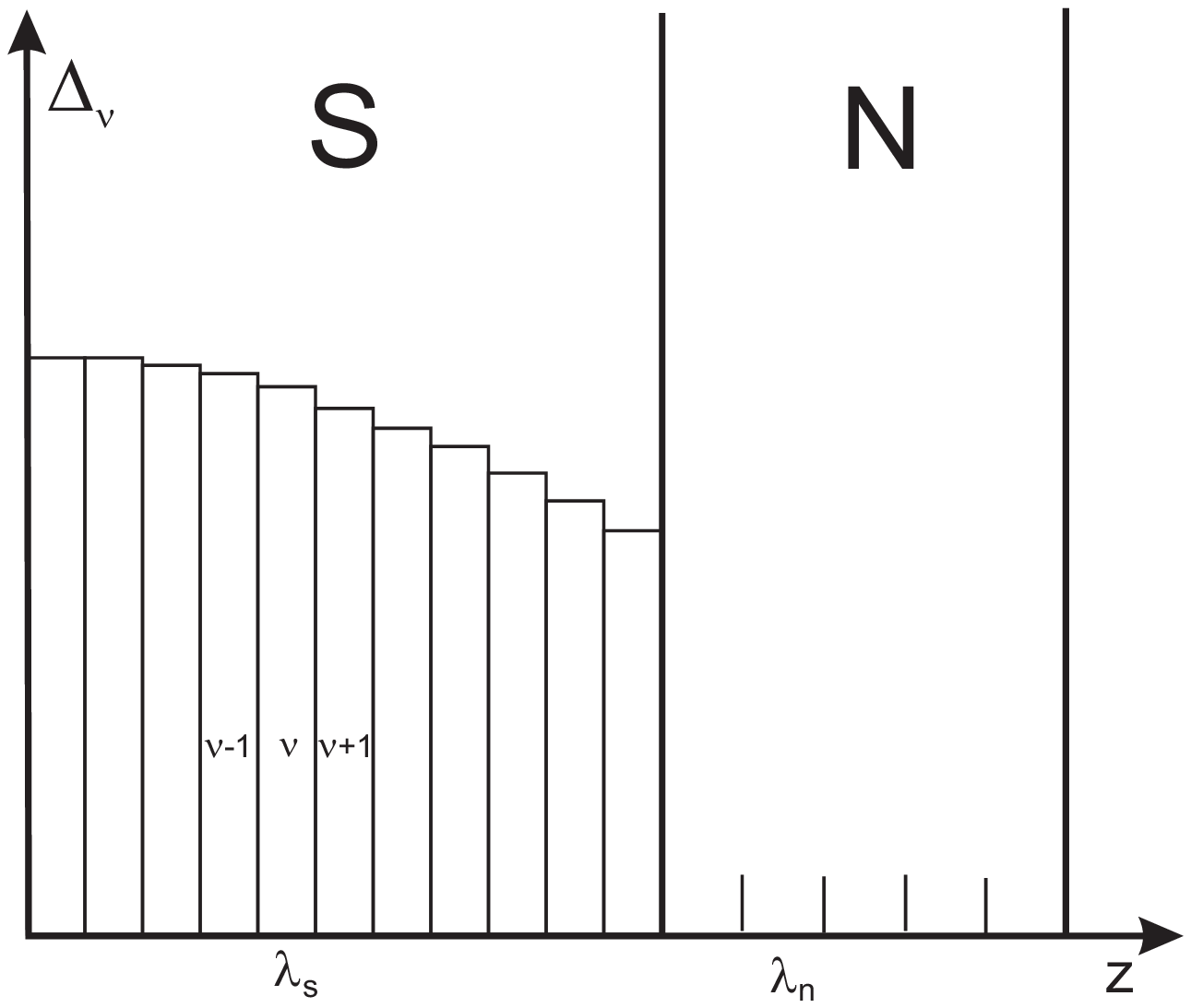}%
\\
Fig.2: The double layer is sliced into sheets of thickness $\lambda
_s$,$\lambda_n$ parallel to the film planes.
\end{center}}}%
\]

\begin{itemize}
\item The superconductor is divided into $Z_{s}$ layers of thickness
$\lambda_{s}$ where $\lambda_{s}=d_{s}/Z_{s}$ ($d_{s}$ is the thickness of the
superconducting film).

\item The BCS interaction $V_{s}$ for the superconductor(s) is fitted, using
the density of states $N_{s}$ and the Debye temperature $\Theta_{D}$(appendix 6.2.1).

\item The time interval $\tau_{d}$ $=2\lambda_{s}/v_{F,s}$ is the time step of
the numerical calculation ($v_{F,s}$ is the Fermi velocity of the
superconductor) (appendix 6.3).

\item For the normal conductor (superconductor with lower $T_{c}$) the same
time step is used by dividing its thickness in layers of thickness
$\lambda_{n}=v_{F,n}\tau_{d}/2$.

\item An initial gap function $\Delta_{\nu}=\Delta\left(  z_{\nu}\right)  $ is
introduced. Each cell is occupied at the time $t^{\prime}=0$ with $O_{\nu
}\left(  0\right)  =N\left(  z_{\nu}\right)  \lambda_{\nu}\Delta\left(
z_{\nu}\right)  $ electrons. ($N\left(  z_{\nu}\right)  $ is the local density
of states, i.e. equal to $N_{s}$ in the superconductor) (appendix 6.2.2).

\item A procedure for diffusive and ballistic propagation of electrons in the
different films is derived (appendix 6.3).

\item The maximal transmission of an electron through the interface in each
direction is calculated. It can be scaled down to include a barrier at the
interface (appendix 6.4).

\item The density $O_{\nu}\left(  m\right)  $ is calculated in descrete steps
for the time $\ t^{\prime}=m\tau_{d}$. (appendix 6.3).

\item Due to thermal dephasing this density is, at each step, multiplied with
the time factor $\eta_{T}\left(  m\tau_{d}\right)  .$

\item The sum $\sum_{m}O_{\nu}\left(  m\right)  \eta_{T}\left(  m\tau
_{d}\right)  $ is formed, multiplied with $\left(  \tau_{d}/\tau_{T}\right)
$/$\lambda_{\nu}$ and, in the superconductor(s), multiplied with $V_{s}$, the
attractive electron-electron interaction.

\item The resulting function $\widetilde{\Delta}_{\nu}$ is the input
$\Delta_{\nu}$ for the next iteration.

\item Since the eigen value has to be 1 the ratio $r=\sum_{\nu}\widetilde
{\Delta}\left(  z_{\nu}\right)  /\sum_{\nu}\Delta\left(  z_{\nu}\right)  $ is
calculated. If $r>1$ $\left(  r<1\right)  $ one increases (lowers) the temperature.

\item The interation process is completed when initial and final $\Delta_{\nu
}$ agree with a relative accuracy of $10^{-5}$. This is generally achieved
after a few iterations.
\end{itemize}

All the step of the numrical procedure are described in details in the appendix.

\newpage

\section{Results}

There are numerous parameters in the superconducting proximity effect: the
coherence lengths $\xi_{s,n}=v_{F}\tau_{T}$ (for the superconductor this is
the BCS $\xi_{BCS}$ if one uses the transition temperature in $\tau_{T}%
=\hbar/\left(  2\pi k_{B}T\right)  $, the mean free path $l_{s,n}$ and the
film thickness $d_{s,n}$ for each film. In addition one has the interface and
the boundaries. Any barrier between the two metals will reduce the transfer
through the interface. Furthermore one can have additional scattering at the
interface between the two films due to a mismatch of the two lattices. The two
surfaces with the vacuum can reflect or scatter the incident electrons or
anything in between. All these scattering parameters influcence the
propagation of the electrons and therefore the transition temperature of the
double layer. In the numerical calculation all these parameters can be
included if they are known or used as fit parameters.

\subsection{Transition temperature}

In the majority of experiments the onset of superconductivity is measured for
a double layer of a thick normal conducting film which is covered with a
superconducting film of increasing thickness. Therefore the first plotted
numerical result represents a double layer of an infinitely thick normal
conductor which is covered with a superconductor of increasing thickness.
Among the large number of possible parameters the following choice is made:
(i) the electronic properties ($N_{s,n}$,v$_{Fs,n})$ of the normal metal and
the superconductor are identical, (ii) the mean free path of the normal
conductor is infinite, (iii) the thickness of the normal conductor is
infinite, (iv) the interface is perfectly transparent, (v) for the mean free
path of the superconductor the following values are chosen: $l_{s}=\infty,$
$\xi_{0}$, $\xi_{0}/10,\xi_{0}/100$. The results are shown in Fig.3a. The
parameter $\alpha$ is defined as $\alpha=l_{s}/\xi_{0}$. The curves of the
transition temperature versus thickness of the superconductor show the typical
behavior; they approach $T_{s}$ for large $d_{s}$ and show a steep decline at
a critical thickness $d_{cr}$. The value of the critical thickness decreases
strongly with decreasing mean free path $l_{s}$ of the superconductor. It
might be surprising that even a mean free path $l_{s}=\xi_{0}$ shifts the
$T_{c}$-$d_{s}$ curve considerably to smaller thicknesses. For the smallest
mean free path of $l_{s}=\xi/100$ the critical thickness is about
$d_{cr}\thickapprox0.19\xi_{0}$. In Fig.3b $T_{c}$ is plotted versus the
reduced thickness $d_{s}/d_{cr}$. The points lie almost on an universal curve,
particularly those for smaller $l_{s}$.
\[%
\begin{array}
[c]{cc}%
{\parbox[b]{3.2777in}{\begin{center}
\includegraphics[
height=2.4259in,
width=3.2777in
]%
{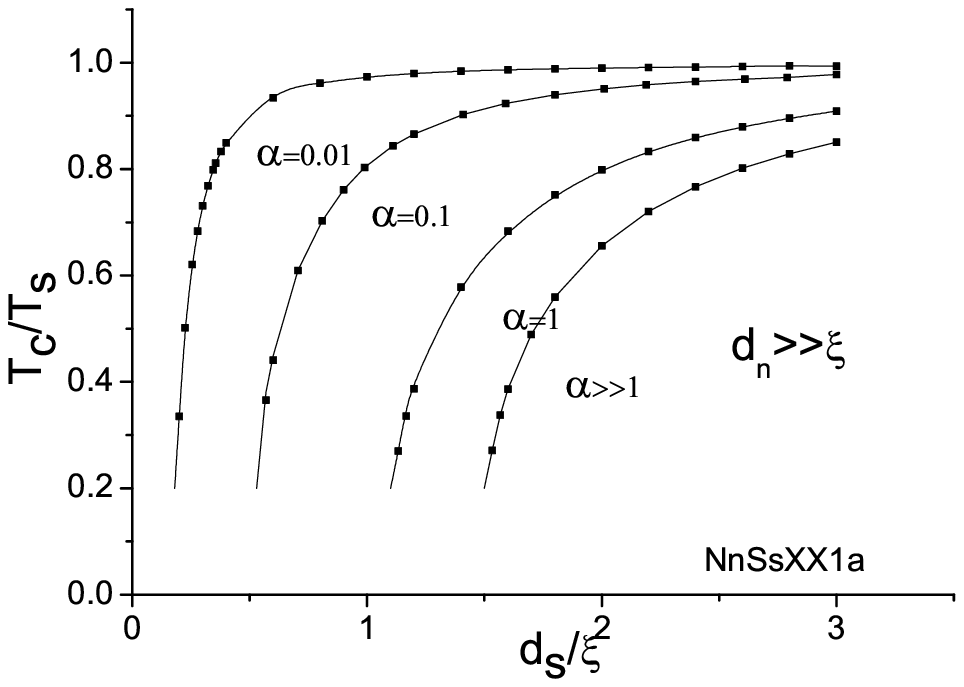}%
\\
Fig.3: The reduced transition temperature $T_c/T_s$ for an NS double layer
where $d_n=\infty$, $l_n=\infty$ as a function of thickness $d_s/\xi$ ($%
\xi$=BCS coherence of S) a) For different mean free paths $l_n$ of S, the
parameter $\alpha=l_s/\xi$.
\end{center}}}%
&
{\parbox[b]{3.0585in}{\begin{center}
\includegraphics[
height=2.347in,
width=3.0585in
]%
{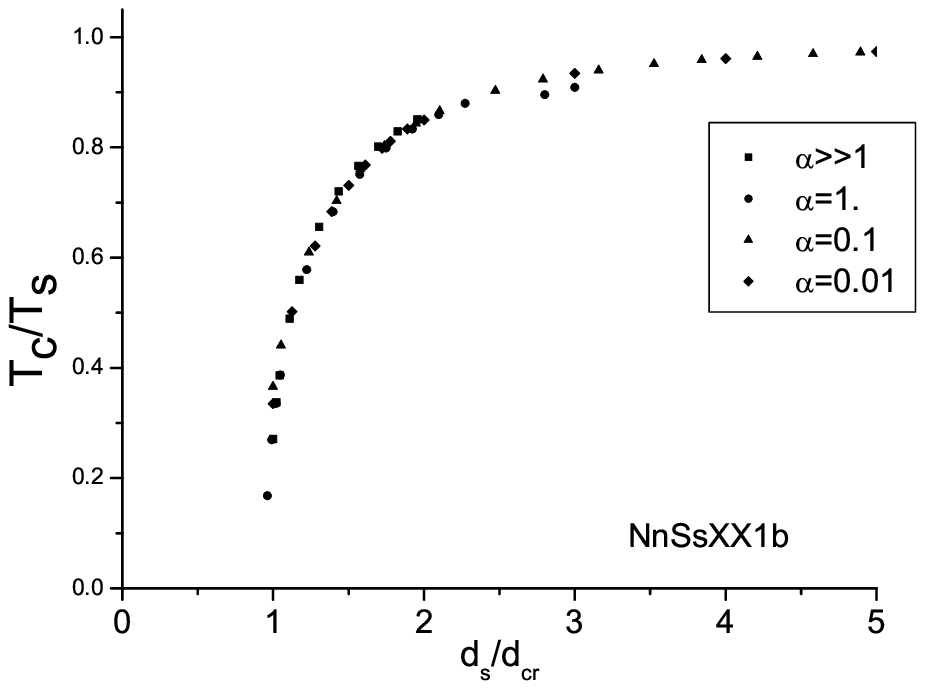}%
\\
Fig.3b: The same plot as a) with the S thickness scaled with the critical
thickness $d_cr.$
\end{center}}}%
\end{array}
\]

\subsection{Pair amplitude}

In the next step the actual dependence of the gap function on position is of
interest. In Fig.4a-d this gap function $\Delta\left(  z^{\prime}\right)  $ is
plotted as a function of $z^{\prime}=z/d_{s}$. We choose double layers where
$T_{c}$ lies in the steep decline of the $T_{c}$ curves in Fig.3 at about
$T_{c}/T_{s}\thickapprox0.3$. Fig.4a shows $\Delta\left(  z^{\prime}\right)
/\Delta_{0}$ for the superconductor with $l_{s}/\xi_{0}=0.01.$ (Since the
amplitude of $\Delta\left(  z^{\prime}\right)  $ approaches zero at the
transition temperature the value $\Delta_{0}$ at the maximum is of no
phyisical significance). Since the gap function has a horizontal slope at the
free surface a comparison with a cosine function $\cos\left(  p\left(
1-z^{\prime}\right)  \right)  $ is useful. The resulting fits are shown in
Fig.4a-d. Fig.4a for $l_{s}/\xi_{0}=0.01$ shows an almost perfect quarter of a
cosine function with $p=1.57$ which is as close to $\pi/2$ as it can be. For
$l_{s}/\xi_{0}=0.1$ the shape of the gap curve is still quite close to a
cosine function but the factor has the value $p=1.46$. For $l_{s}/\xi_{0}=1$
the shape of the gap curve shows already clear deviations from a cosine curve
and the coefficient is $p\thickapprox1.25$. Finally in the clean limit the gap
function curves stronger for small $z^{\prime}$ than the cosine curve and the
coefficient is $p\thickapprox1.05$ for the shown fit. This behavior is
interesting because in a number of theoretical papers the gap function is
expanded into a series (see for example ref. \cite{F34} where a series
consisting of $\cos\left(  \Omega_{0}\left(  d_{s}-z)/\xi_{d}\right)  \right)
$ and $\cosh\left(  \left(  \Omega_{m}\left(  z-d_{s}\right)  /\xi_{d}\right)
\right)  $ is used, $\xi_{d}=\sqrt{l_{s}\xi} $ is the superconducting
diffusion length and $\Omega_{0}$,$\Omega_{m}$ are coefficients defined in
that work).%

\[%
\begin{array}
[c]{cc}%
{\includegraphics[
height=2.2283in,
width=2.6642in
]%
{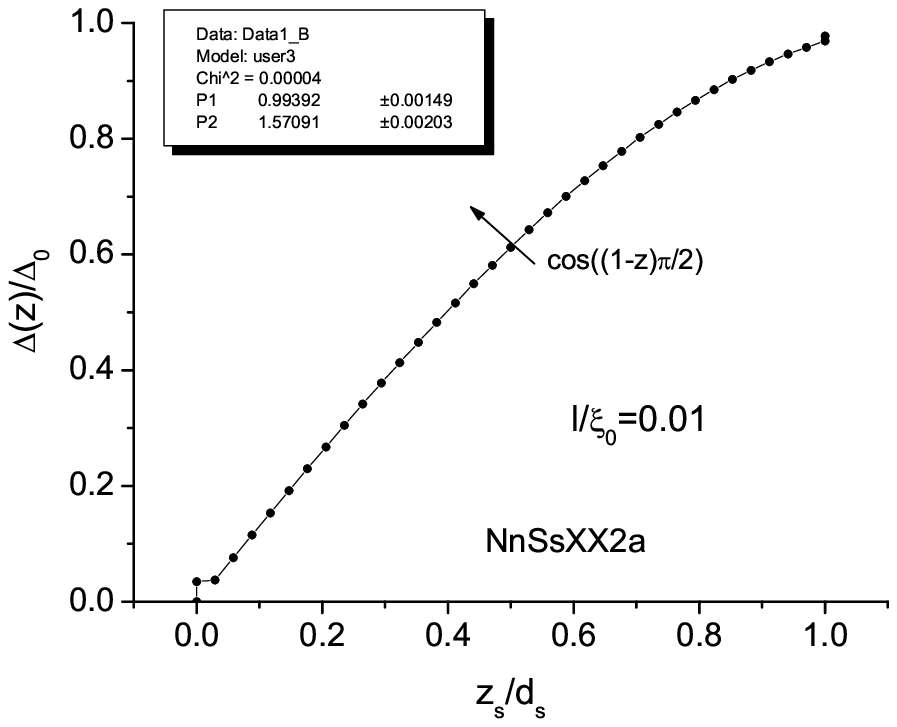}%
}%
&
{\includegraphics[
height=2.0988in,
width=2.8667in
]%
{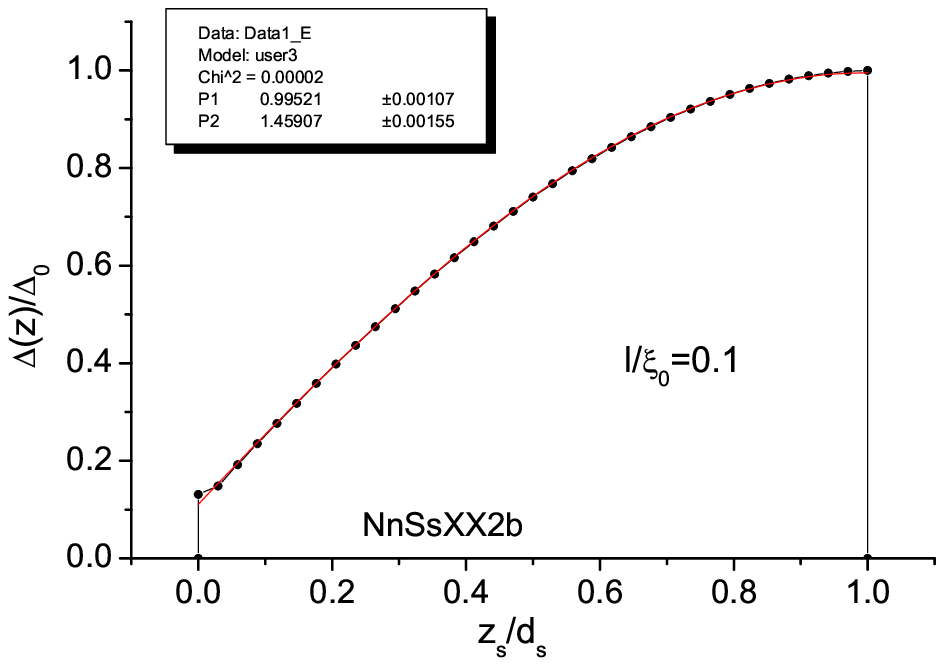}%
}%
\\%
{\includegraphics[
height=2.2806in,
width=2.7729in
]%
{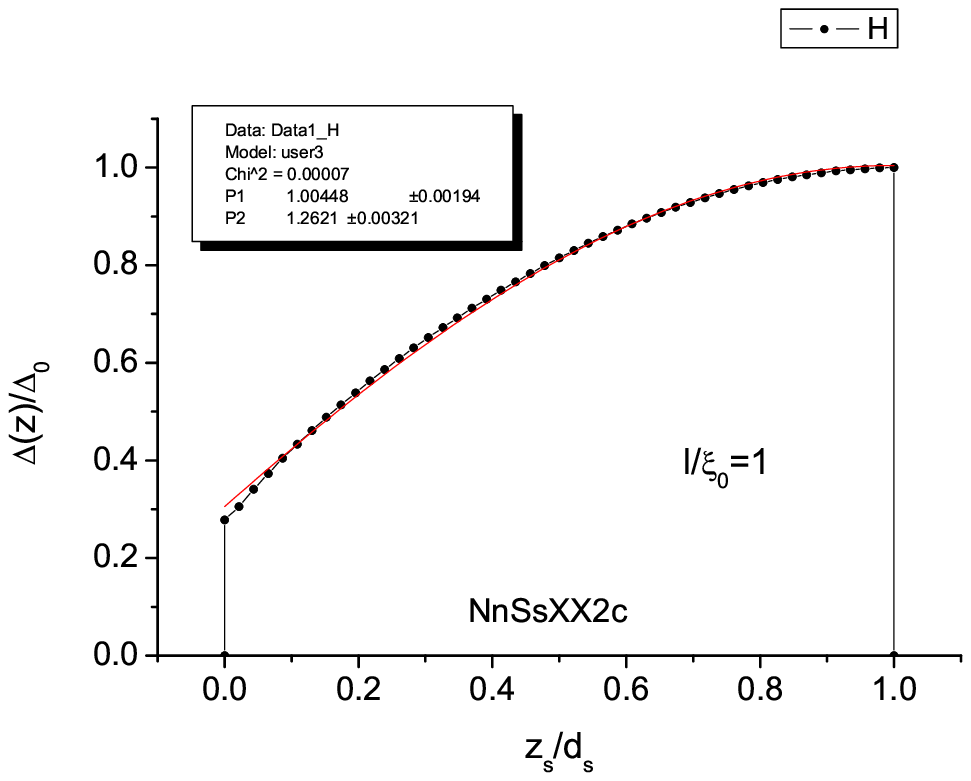}%
}%
&
{\includegraphics[
height=2.0108in,
width=2.851in
]%
{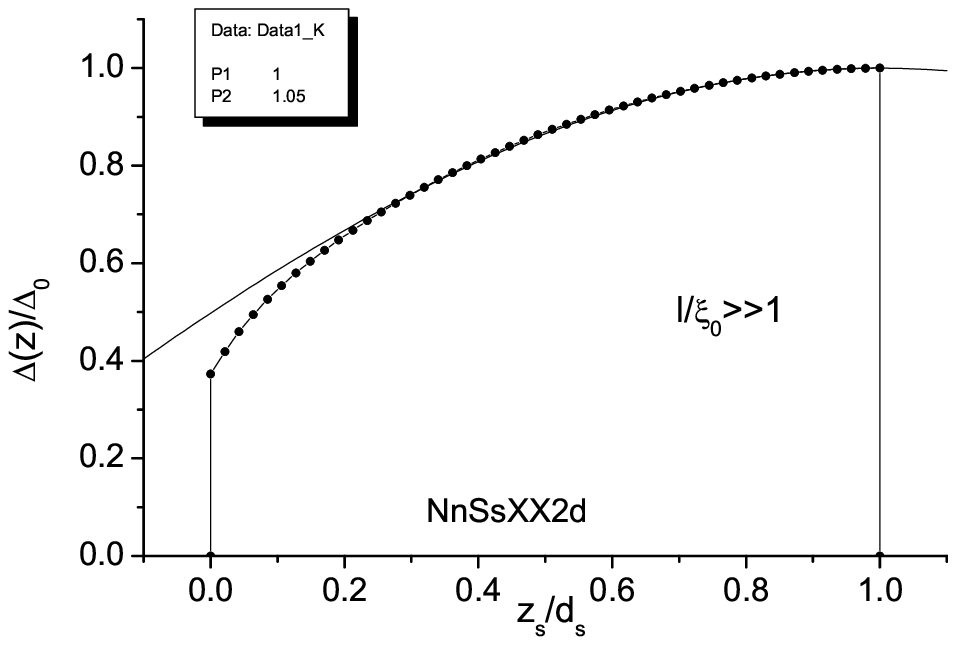}%
}%
\end{array}
\]
$%
\begin{tabular}
[c]{l}%
Fig.4a-d: The gap function $\Delta\left(  z\right)  $ is plotted versus the
position $z/d_{s}$ in the\\
superconductor for NS double layers. Each drawing corresponds to one\\
of the curves in Fig.3a. $d_{s}$ is close to the critical thickness $d_{cr}$.
The ratio\\
$l_{s}/\xi_{0}$ is noted in the figures.
\end{tabular}
$

The simple form of the gap function in the case of $l_{s}/\xi_{0}=0.01$ makes
it very obvious why the very disordered superconductors (often
discriminatingly called dirty superconductors) are much easier to describe.
This becomes still more obvious if one compares the shape of the gap function
at different $T_{c}/T_{s}$ values (which means, of course, using different
thicknesses of the superconductor). In Fig.5a the (normalized) gap functions
for $T_{c}/T_{s}$ \ values of about 0.3 and 0.9 are shown as a function of
$z/d_{s}$. They lie perfectly on the same quarter of a cosine function. This
is very different for the clean limit where the shape depends strongly on the
temperature.%
\[%
\begin{array}
[c]{cc}%
{\parbox[b]{2.8427in}{\begin{center}
\includegraphics[
height=2.2665in,
width=2.8427in
]%
{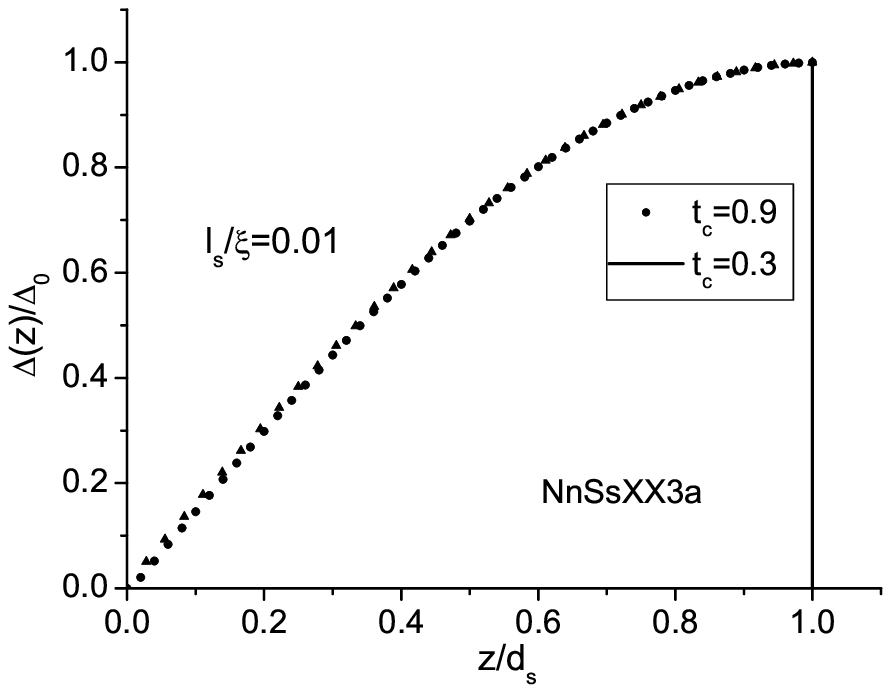}%
\\
Fig.5a,b: The gap function $\Delta\left(  z\right)  $ is plotted versus the
position $z/d_s$ for two NS double layers, each at two different $d_s$
(resulting in different T$_c$ of about 0.3 and 0.9 $T_s$.) a) dirty limit
$l_s/\xi$=0.01.
\end{center}}}%
&
{\parbox[b]{2.831in}{\begin{center}
\includegraphics[
height=2.2839in,
width=2.831in
]%
{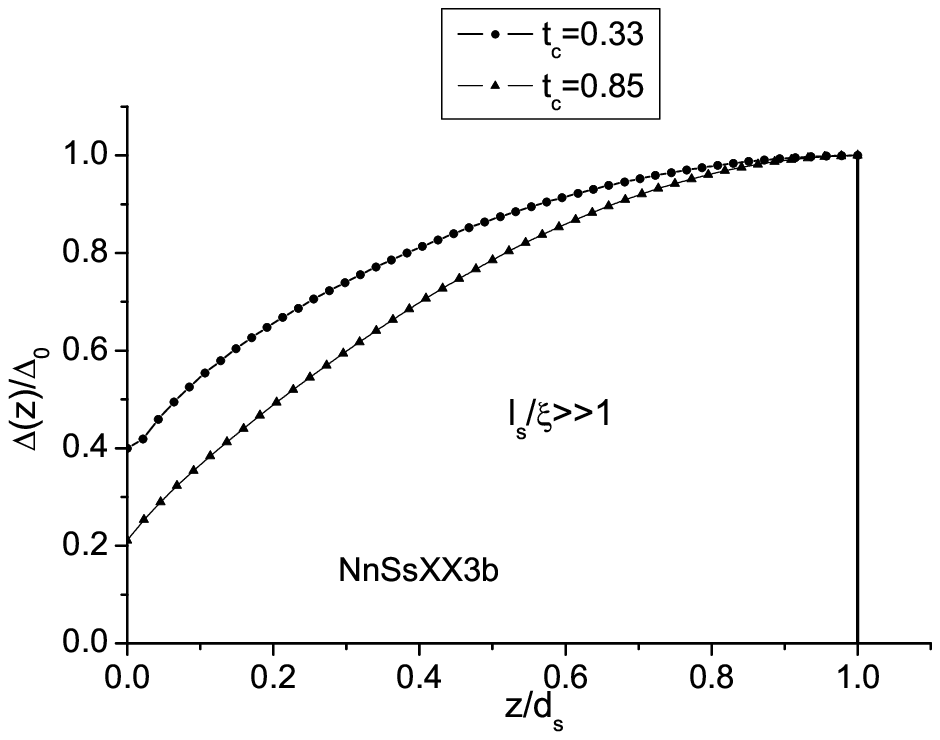}%
\\
b) clean limit $l_s=\infty$.
\end{center}}}%
\end{array}
\]

\subsection{Dirty limit}

Since in the dirty limit the gap function approached such a simple form for a
superconductor in contact with an infinite clean normal metal it is worth to
check the situation when both metals are dirty. This is the case which most
theoretical papers investigate.
\[%
\begin{array}
[c]{cc}%
{\parbox[b]{3.2453in}{\begin{center}
\includegraphics[
height=2.6227in,
width=3.2453in
]%
{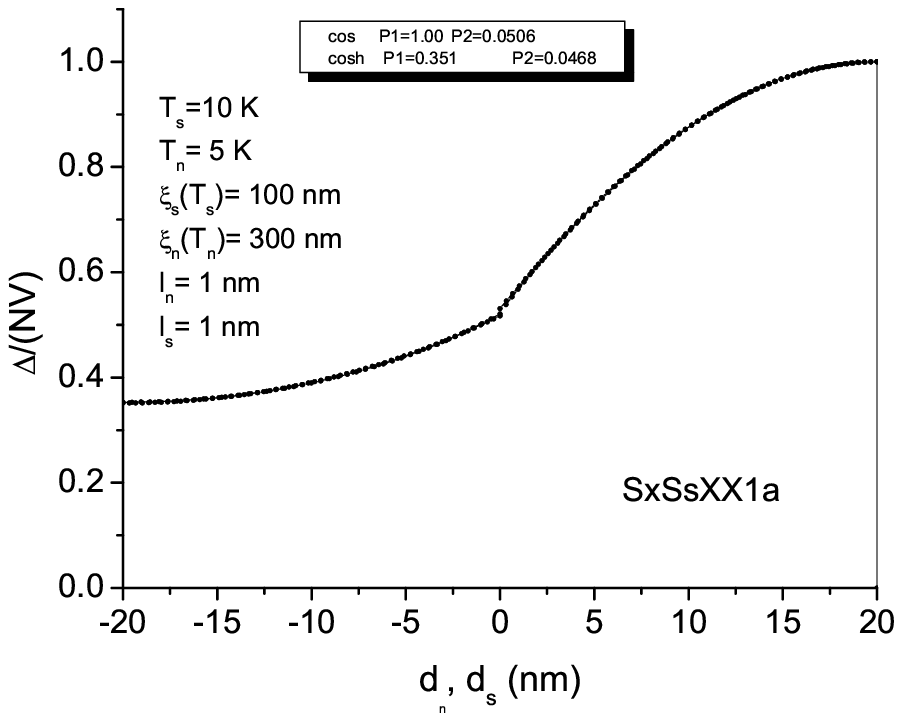}%
\\
Fig.6a,b: The function $\Delta/\left(  NV\right)  $ for an S$_\text{1}%
$S$_\text{2}$ double layer, $T_c1=T_s2/2$ (details in text and figures). Both
films are in the dirty limit. text and figures). Both films are in the dirty
limit. In a) the transmission coefficient from S$_2$ to S$_1$ is $t=1.0$ whil
in b) $t=0.8 $.
\end{center}}}%
&
{\parbox[b]{3.1656in}{\begin{center}
\includegraphics[
height=2.8269in,
width=3.1656in
]%
{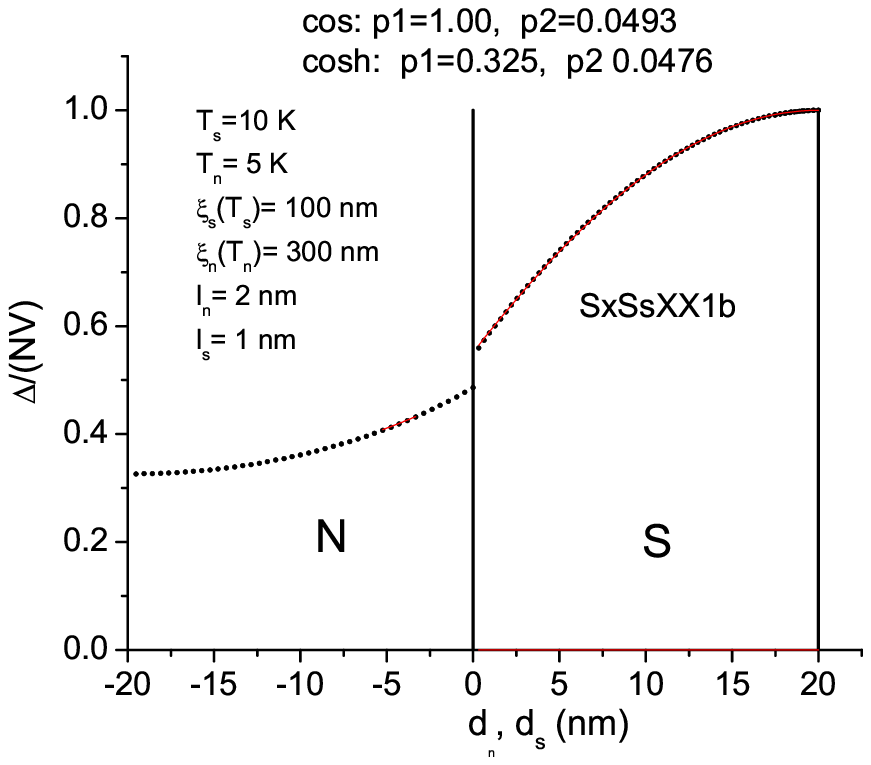}%
\end{center}}}%
\end{array}
\]

In Fig.6a $\Delta/\left(  NV\right)  $ is plotted for a double layer of two
superconductors with different transition temperatures of $T_{n}=5K$ and
$T_{s}=10K$. In addition the density of states for superconductor N (with the
lower transition temperature) is larger by a factor of 1.5 than for
superconductor S. Therefore the superconducting coherence lengths $\xi
_{0s}=\hbar v_{Fs}/\left(  2\pi k_{B}T_{s}\right)  =100nm$ and $\xi_{0n}=\hbar
v_{Fn}/\left(  2\pi k_{B}T_{n}\right)  =300nm$ are different (The additional
factor of two stems from the ratio of the transition temperatures). The
difference in the density of states and the Fermi velocity of the two metals
yields a ratio of the two transmission coefficients at the interface
$T_{N->S}/T_{S->N}=.444$.

The thickness of each film is $d_{n}=d_{s}=20nm$. In Fig.6a the mean free
paths are chosen in both films to be $l_{s}=l_{n}=1nm$. For the corresponding
superconducting diffusion lengths $\xi_{ds},\xi_{dn}$ one finds $\xi
_{ds}=\sqrt{\xi_{0s}l_{s}}=10nm$ and $\xi_{dn}=\sqrt{\xi_{0n}l_{n}%
}=17.\,\allowbreak3nm$. According to de Gennes the function $\Delta/\left(
NV\right)  $ should be continuous at the interface. As can be easily
recognized from the plot in Fig.6a this condition is well fulfilled. Werthamer
\cite{W32} expressed the z dependence of the gap function $\Delta\left(
z\right)  $ in the two superconductors as
\[%
\begin{tabular}
[c]{lll}%
$\cosh\left(  k_{n}\left(  d_{n}+z\right)  \right)  $ & in the superconductor
with $T_{n}$ in the range $-d_{n}<z<0$ & \\
$\cos\left(  k_{s}\left(  d_{s}-z\right)  \right)  $ & in the superconductor
with $T_{s}$ in the range $0<z<d_{s}$ &
\end{tabular}
\]
Fig.6a shows for $z<0$ a fit to the function $a\cosh\left(  k_{n}\left(
d_{n}+z\right)  \right)  $ and for $z>0$ to the function $\cos\left(
k_{s}\left(  d_{s}-z\right)  \right)  $. The fitted curves lie within the
trace of the points. The fitted values for the parameters are $k_{s}%
=0.0506nm^{-1}$, $k_{n}=0.0468nm^{-1}$ and $a=0.351$. This yields for the
value of $\Delta/\left(  NV\right)  $ on the left and the right side of the
interface: $0.516$ and $0.530$. The corresponding slopes on the left and right
side of the interface are: $4.\,\allowbreak29\times10^{-2}$ and
$1.\,\allowbreak77\times10^{-2}$. According to de Gennes the derivative
$\left(  D/V\right)  d\Delta/dz$ should be continuous at the interface for the
dirty limit. Using the input data of the two superconductors $D_{s,n}$ and
$V_{s,n}$ one obtains for the ratio of the slopes $2.\,\allowbreak61$. The
simulated $\Delta\left(  z\right)  $ yields a slope ratio at the interface of
$2.\,\allowbreak42$. So the de Gennes condition is verified with an accuracy
of about 10\%.

In a second simulation the transmission through the interface is reduced by a
factor 2. It is quite remarkable that this changes the transition temperature
only from $T_{c}=7.6K$ by just $0.1K$ to $7.7K$. In Fig.6b the function
$\Delta/\left(  NV\right)  $ is plotted for the double layer as a function of
$z$. One recognizes that now $\Delta/\left(  NV\right)  $ is no longer
continuous at the interface. The functional form in N and S can still be well
fitted by a hyperbolic cosine and a cosine function. (The fitted curves lie
within the width of the numercal points).

\subsection{Initial slope}

When one condenses the normal metal on top of the superconductor then the
transition temperature of the double layer decreases. Here the focus is on the
question how the intial slope at $d_{n}=0$ depends on various parameters, such
as the mean free path in the superconductor and the normal conductor and the
transparency of the interface.

The dependence of the initial slope on the mean free path is shown in
Fig.7a,b. In both figures the thickness of the superconductor is equal to the
BCS coherence length $\xi_{0}.$The transition temperature $T_{c}/T_{s}$ is
plotted versus the thickness of the normal conductor. In Fig.7a the mean free
paths in both films are equal and vary between $l_{s}=l_{n}=\xi_{0}/10, $
$\xi_{0}$ and $10^{3}\xi_{0}$. In Fig.7b four different combinations of
$\left(  l_{s},l_{n}\right)  $ are chosen. From the top to the bottom $\left(
l_{s},l_{n}\right)  $ is equal to $\left(  \xi_{0}/100,\xi_{0}/100\right)  $,
$\left(  \xi_{0}/100,10^{3}\xi_{0}\right)  $, $\left(  10^{3}\xi_{0},\xi
_{0}/100\right)  $ and $\left(  10^{3}\xi_{0},10^{3}\xi_{0}\right)  $. For all
curves the initial slope is identical. (In all the numerical calculations
which were discussed so far the two density of states are assumed equal
$N_{s}=N_{n}$.)%

\[%
\begin{array}
[c]{cc}%
{\parbox[b]{3.0959in}{\begin{center}
\includegraphics[
height=2.2524in,
width=3.0959in
]%
{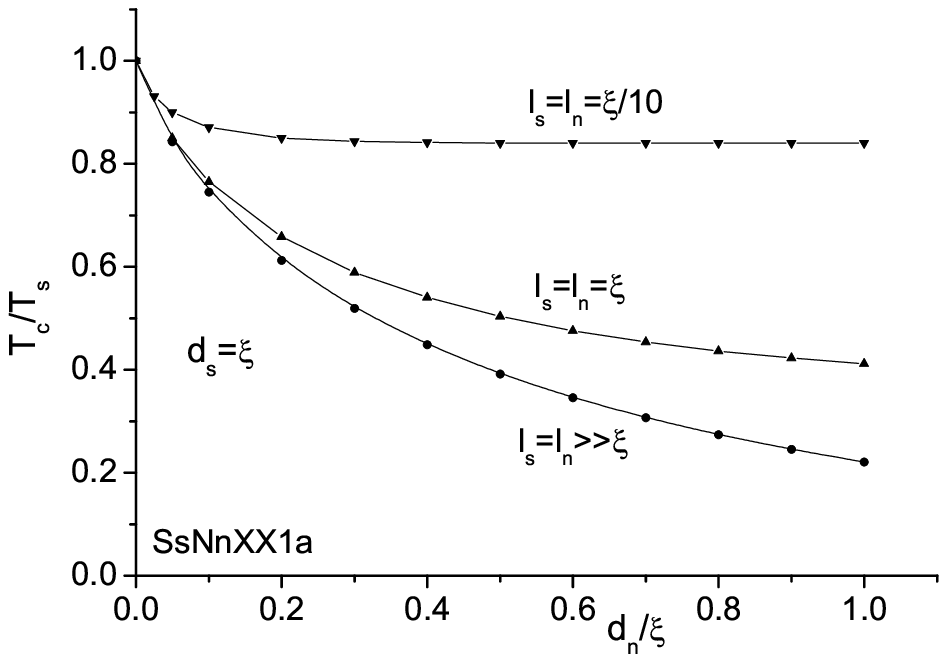}%
\\
Fig.7a,b: $T_c$ for an SN double layers as a function of $d_n/\xi$. The
thickness of the superconductor is equal to the BCS coherence length $\xi$.
($N_s=N_n$). a) The mean free paths $l_s=l_n$ are parameters.
\end{center}}}%
&
{\parbox[b]{2.7314in}{\begin{center}
\includegraphics[
height=2.2698in,
width=2.7314in
]%
{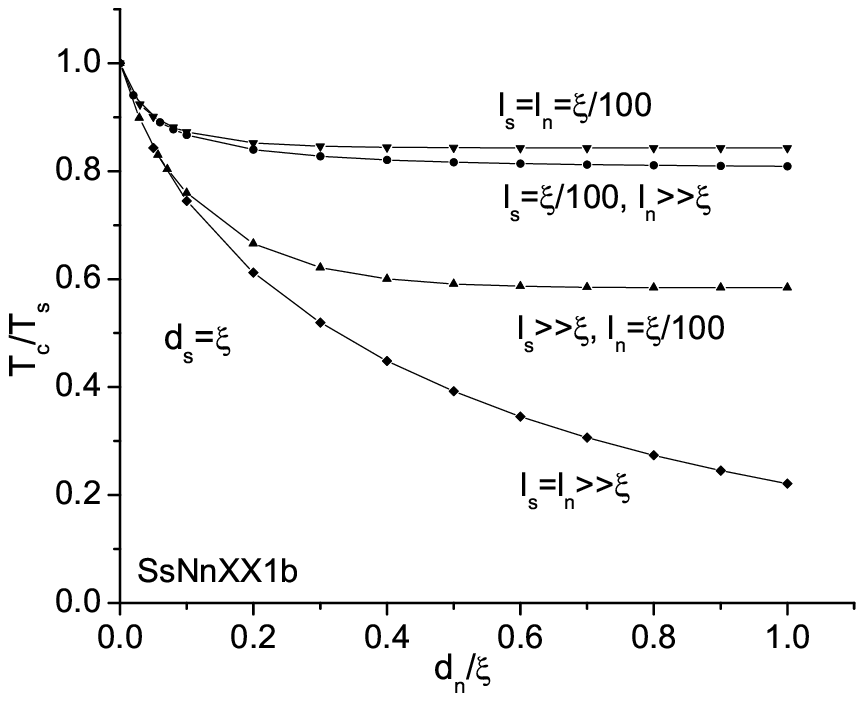}%
\\
b) Different combinations of the mean free paths are used as parameters.
\end{center}}}%
\end{array}
\]

In Fig.8 the dependence of the initial slope on the thickness of the
superconducting first layer is tested. The graph shows the dependence of
$T_{c}/T_{s}$ for a small range of the thickness $d_{n}$ of the normal
conductor to emphasize the initial range.%

\[%
{\parbox[b]{3.1474in}{\begin{center}
\includegraphics[
height=2.5122in,
width=3.1474in
]%
{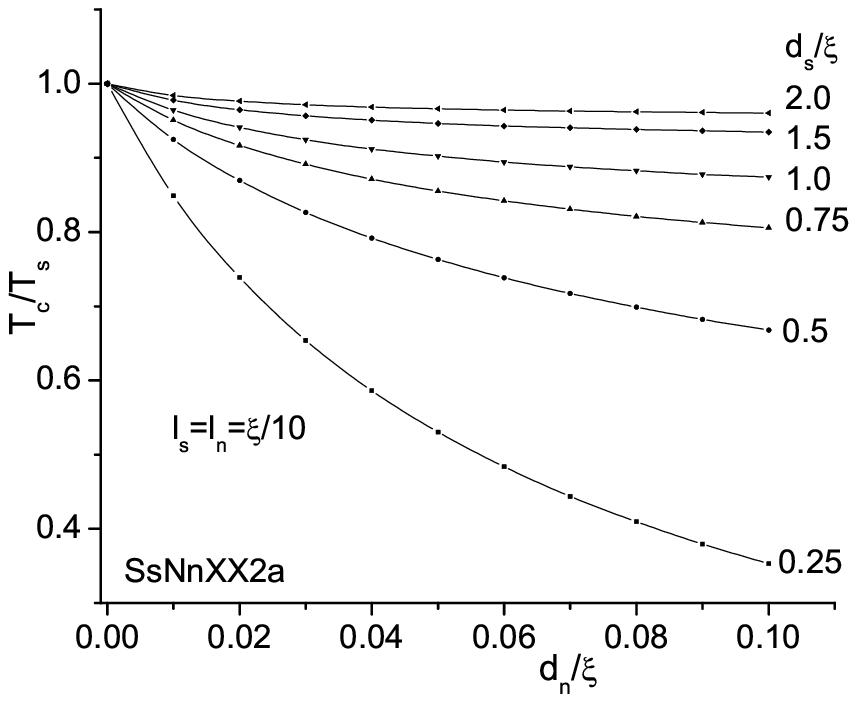}%
\\
Fig.8: $T_c$ for SN double layers as a function of $d_n/\xi$. The parameter
$d_s$ is the thickness of the superconductor$.$ ($N_s=N_n$, $l_s=l_n=\xi/10$).
\end{center}}}%
\]
In table I the normalized initial slope is collected. (The numerical points
had to fitted with a polynomial to extract the slope from the numerical
results). Up to a thickness of $d_{s}=\xi$ the $S_{sn}$ is constant within
about $\pm1\%$. For larger $d_{s}$ it decreases slightly. But since the value
of $dT_{c}/dd_{n}$ becomes quite small this thickness range is not well suited
for the experimental determination of the slope. The main result is that the
normalized initial slope is essentially independent of the thickness of the
superconductor.
\begin{align*}
&
\begin{tabular}
[c]{|l|l|}\hline
$\dfrac{\mathbf{d}_{s}}{\xi}$ & $\dfrac{d_{s}}{T_{s}}\dfrac{dT_{c}}{dd_{n}}%
$\\\hline
0.25 & 4.35\\\hline
$0.5$ & 4.34\\\hline
0.75 & 4.31\\\hline
1.0 & 4.26\\\hline
1.5 & 4.13\\\hline
2.0 & 4.02\\\hline
\end{tabular}
\\
&
\begin{tabular}
[c]{l}%
Table I: Normalized initial slope for different\\
thicknesses $d_{s}$ of the superconductor
\end{tabular}
\end{align*}

Finally Fig.9 shows that the initial slope does not depend on the transmission
through the interface. In this calculation the density of states in both
metals is chosen to be equal $N_{s}=N_{n}$ and the mean free paths are
$l_{s}=l_{n}=\xi/10$. The transmission coefficient is varied between 0.2 and
1.0. The resulting $T_{c}$-$d_{n}$ curves show the same initial slope.%

\[%
{\parbox[b]{3.2088in}{\begin{center}
\includegraphics[
height=2.7048in,
width=3.2088in
]%
{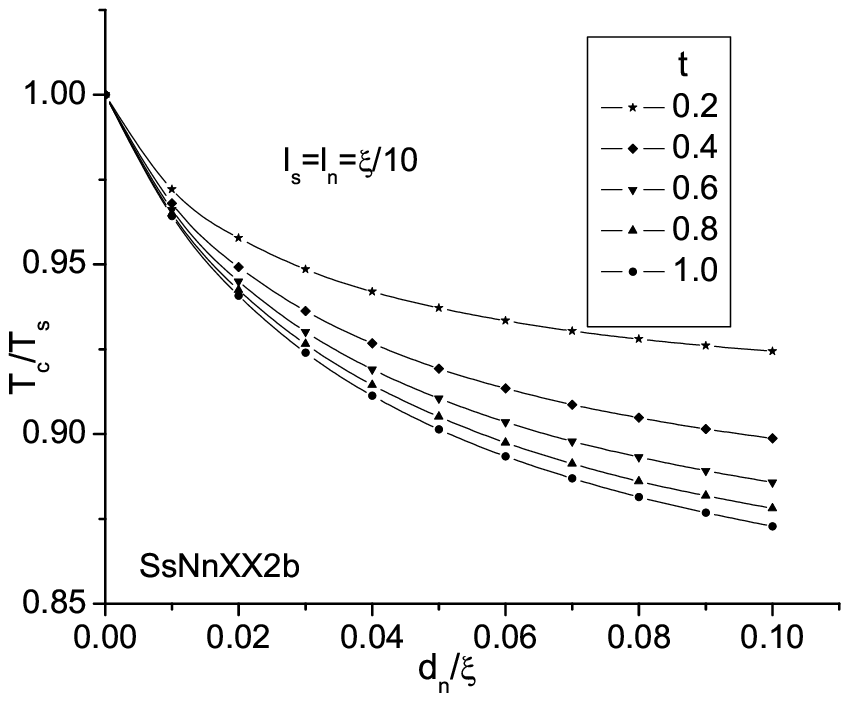}%
\\
Fig.9: $T_c$ for an SN double layers as a function of $d_n/\xi$. The parameter
$t$ is the transparency of the interface$.$($N_s=N_n,$ $d_s=\xi$, $%
l_s=l_n=\xi/10$).
\end{center}}}%
\]

\newpage

\section{Discussion}

The intention of this paper was to develop a convenient numerical procedure
for the superconducting proximity effect so that graduate students could
instantly compare their experimental results with the theory. One importat
result of this investigation is the fact that the (normalized) initial slope
of an SN double layer is independent of most film parameters except the
density of states ratio and the effective BCS interaction.
\begin{equation}
S_{sn}=\frac{d_{s}}{T_{c0}}\left\vert \frac{dT_{c}}{dd_{n}}\right\vert
=\Gamma_{sn}\frac{N_{n}}{N_{s}}\label{slp}%
\end{equation}

In the case of a weak coupling superconductor $\Gamma_{sn}$ is given by the
Cooper limit, i.e. $\Gamma_{sn}=1/\left(  NV\right)  _{s}$, the inverse of the
BCS interaction. If the Debye temperature is not several orders of magnitude
larger than $T_{s}$ then one has to determine $\Gamma_{sn}$ in equation
(\ref{slp}) numerically. Using $T_{s}=7.2K\ $\ for Pb then the prefactor is
about 4.5. (This is actually the value for a wide range of the Debye
temperature between $100K$ and $300K$).

Recently our group investigated the proximity effect between the
superconductor Pb and several alkali metals \cite{B135}. It was a great
surprise that the experimental initial slope of these SN double layers could
not be explained with the density of states from the literature. Instead the
experimental $\left(  d_{s}/T_{s}\right)  \left(  dT_{c}/dT\right)  $ was too
small by more than a factor of two. Table II gives some of the data of the SN
double layers. (The thickness of the normal metal was the smallest thickness
in a full curve.)%
\begin{align*}
&
\begin{tabular}
[c]{cccccc}%
\textbf{metals} & $\mathbf{d}_{s}$ $(nm)$ & $\mathbf{d}_{n}$ $(nm)$ &
$\mathbf{S}_{sn}|_{\exp}$ & $\mathbf{N}_{\mathbf{n}}\mathbf{/N}_{\mathbf{s}}$
& \textbf{ratio}\\
Pb/K & 12.9 & 2.04 & 0.423 & 0.223 & 1.90\\
Pb/Na & 13.9 & 2.18 & 0.546 & 0.300 & 1.82\\
Pb/Ag & 17.9 & 2.10 & 0.625 & 0.335 & 1.86\\
&  &  &  &  &
\end{tabular}
\\
&
\begin{tabular}
[c]{l}%
Table II: The normalized initial slope of SN double layers with Pb as\\
superconductor and different normal metals. The columns 2-6 give the\\
thickness of the superconductor, the normal conductor, the experimental\\
initial slope, the ratio of the density of states and the ratio $\mathbf{S}%
_{sn}|_{\exp}/\left(  N_{n}/N_{s}\right)  $.
\end{tabular}
\end{align*}

We searched the literature for other measurements of SN double layers and
their initial slope. It turned out that there are very few measurements of SN
layers. (At this stage we excluded transition metals because they show
two-band superconductivity and it is not obvious how the different
superconducting bands couple to the normal conductor). There were essentially
two groups of publications which had measured SN double layers which contained
information about the intial slope. The first group of papers was by Hilsch et
al.\cite{H31}, \cite{H26} who investigated quench condensed PbCu layers. The
second work was by Minnigerode \cite{M45} who also investigated PbCu layers
but prepared the layers at room temperature. Particularly the second paper
gives detailed tables of thicknesses of the two components and transition
temperatures. The results of these papers are collected in table III. The
first column gives the components of the SN double layer, the second and third
columns the thicknesses of the superconductor and normal metal. The fourth
column contains the experimental normalized slope and the fifth gives the
ratio $N_{n}/N_{s}$. The last column contains the ratio of the experimental
(normalized) slope to the density of states ratio. Again the experimental
normalized slopes are much smaller than the theory predicts.%
\[%
\begin{array}
[c]{c}%
\begin{tabular}
[c]{cccccc}%
\textbf{metals} & $\mathbf{d}_{s}$ $(nm)$ & $\mathbf{d}_{n}$ $(nm)$ &
$\mathbf{S}_{sn}|_{\exp}$ & $\mathbf{N}_{\mathbf{n}}\mathbf{/N}_{\mathbf{s}}$
& \textbf{ratio}\\
Pb/Cu$^{1}$ & 10.0 & 10.0 & 0.542 & 0.448 & 1.21\\
Pb/Cu$^{1}$ & 15.0 & 10.0 & 0.500 & 0.448 & 1.12\\
Pb/Cu$^{2}$ & 22.9 & 3.30 & 1.110 & 0.448 & 2.48\\
Pb/Cu$^{2}$ & 24.3 & 3.90 & 0.935 & 0.448 & 2.08\\
Pb/Cu$^{2}$ & 32.9 & 4.10 & 0.883 & 0.448 & 1.97\\
Pb/Cu$^{2}$ & 27.2 & 13.10 & 0.683 & 0.448 & (1.52)\\
Pb/Cu$^{2}$ & 28,0 & 26.40 & 0.842 & 0.448 & (1.88)\\
Pb/Cu$^{2}$ & 33.4 & 17.70 & 0.608 & 0.448 & (1.36)
\end{tabular}
\\%
\begin{tabular}
[c]{l}%
Table III: The normalized initial slope of PbCu double layers, data$^{1}$
from\\
ref. \cite{H31} are quench condensed\ and data$^{2}$ from ref. \cite{M45}
are\\
condensed at room temperature. The columns 2-6 give the thickness of\\
the superconductor, the normal conductor, the experimental initial slope,\\
the ratio of the density of states and the ratio $\mathbf{S}_{sn}|_{\exp
}/\left(  N_{n}/N_{s}\right)  $.
\end{tabular}
\end{array}
\text{\ \ \ }%
\]

It is rather amazing that this fundamental discrepancy between experiment and
theory has not been realized. What is the reason for this disagreement? The
authors best guess at the present time is that the use of the weak coupling
theory of superconductitivity is not adaquate for the double layers containing
the supercondutor Pb. The superconductor Pb is a convenient component of an SN
double layer because it has a rather large $T_{c}$ and is easy to condense.
However, Pb is a strong coupling superconductor. The Fermi sphere of free
electrons is modified by the electron-phonon interaction. An electron
$\mathbf{k}$ which lies below the Fermi energy within the Debye energy
$k_{B}\Theta_{D}$ can emit a virtual phonon and perform a transition into a
state $\mathbf{k}^{\prime}$ above the Fermi energy. As a consequence the
states below $k_{F}$ have an occupation less than 1 and the states above
$k_{F}$ have an occupation larger than 0, even at zero temperature. At $k_{F}
$ the occupation does not jump from one to zero, but has a smaller step of
$1/\left(  1+\lambda\right)  $, where $\left(  1+\lambda\right)  $ is the
electron-phonon enhancement factor. The parameter $\lambda$ can be calculated
with the Eliashberg function $\alpha^{2}F\left(  \Omega\right)  $ which
desribes the probability of an electron to emit or absorb a phonon of energy
$\hbar\Omega$. (A nice treatment of the electron-phonon physics can by found
in the book by Grimvall \cite{G42}). If one moves now an electron from the
Fermi energy to a k-value above the Fermi energy then it changes the
occupation of the k-state only by $\left(  1+\lambda\right)  ^{-1}$. The
change in kinetic energy is therefore reduced by the same factor; the energy
states lie closer together by the factor $\left(  1+\lambda\right)  $. As a
consequence the density of states is enhanced by $\left(  1+\lambda\right)  $
and the Fermi velocity reduced by $1/\left(  1+\lambda\right)  $. This
enhancement is restricted to a small region in the vicinity of the Fermi
energy. The new density of states $N^{\ast}=N\left(  1+\lambda\right)
=N=m\left(  1+\lambda\right)  k_{F}/\left(  2\pi^{2}\hbar^{2}\right)  $ and
Fermi velocity $v_{F}^{\ast}=v_{F}/\left(  1+\lambda\right)  =\hbar
k_{F}/m\left(  1+\lambda\right)  $ are called the renormalized density of
states and Fermi velocity and marked with a star. The Fermi wave number
$k_{F}$ does not change and both changes in $N$ and $v_{F}$ can be expressed
by a change of the mass $m^{\ast}=m\left(  1+\lambda\right)  $ (therefore the
name mass enhancement) .

Many properties of the electron gas and the strong coupling superconductor can
be reasonably well calculated in a renormalized strong coupling description.
For the superconductivity this means that one applies the weak coupling BCS
theory using renormalized density of states and Fermi velocity. The error for
Pb is generally less than 20\% as the ratio of $2\Delta_{0}/k_{B}T_{c}=4.2$
instead of 3.5 demonstrates. (For other parameters see \cite{B44}).

In strong coupling superconductors the effective interaction $\left(
NV\right)  _{s}$ is replaced by $\left(  \lambda-\mu^{\ast}\right)  $ where
$\lambda$ describes the strength of the electron-phonon interaction and
$\mu^{\ast}$ is the Coulomb pseudo-potential. In the literature one find
$\lambda$-values for Pb which vary between 0.8 and \ 1.6 \cite{G42} .
Depending how much the strong coupling theory modifies the expression for the
initial slope it could alter dramatically the theoretical prediction for the
initial slope. In the appendix we show that a renormalized strong coupling
theory would yield an initial slope of $\Gamma_{sn}=\left(  \lambda-\mu^{\ast
}\right)  /\left(  1+\lambda\right)  $. This appears to alter the theoretical
value for the initial slope. The difficulty is that the weak coupling
treatment just a replaces $\left(  NV\right)  _{s}$ by $\left(  \lambda
-\mu^{\ast}\right)  /\left(  1+\lambda\right)  $. This means that for the
transition temperature one obtains the equivalent condition
\[
\Delta=\dfrac{2\pi k_{B}T}{\hbar}\sum_{\left\vert \omega_{j}\right\vert
\leq\Omega_{D}}\dfrac{\lambda-\mu^{\ast}}{1+\lambda}\dfrac{1}{2\left\vert
\omega_{j}\right\vert }\Delta
\]
This does not alter the situation and yields, as observed by Morel and
Anderson \cite{A64}, a too small value of $0.4$ for $\lambda$.

An obvious proposal would be to solve the superconducting proximity effect for
strong coupling superconductors. This means to develop and solve a series
equations for the energy and position dependent gap function $\Delta\left(
\mathbf{r},\omega_{l}\right)  $ which has the equations $($\ref{gap2c}) and
(\ref{scge}) as limiting cases. This is a very demanding job which goes beyond
the scope of the present paper and has to be left for future investigations.
However, we can check whether this extension is a promissing one. For this
purpose we consider the analogy to the Cooper limit for strong superconductors.

Let us consider a double layer composed of (very) thin films of a strong
coupling superconductor and a normal metal. The super- and normal conductor
have $\lambda$ values $\lambda_{s}$ and $\lambda_{n}$. Both have the same
value for the renormalized Coulomb repulsion $\mu^{\ast}$. For sufficiently
thin films the electrons in the double layer travel so quickly from the
superconductor to the normal metal and vice versa that they average over the
properties of the two metals. We have essentially a new superconductor with a
new averaged electron-phonon interaction $\overline{\lambda}$ (in complete
analogy to Cooper's arguement)
\[
\overline{\lambda}=\dfrac{d_{s}N_{s}\lambda_{s}+d_{n}N_{n}\lambda_{n}}%
{d_{s}N_{s}+d_{n}N_{n}}%
\]
Now we apply the strong coupling gap equation (\ref{scge}) to this artificial
strong coupling superconductor.

For the superconductor we chose Pb, but somewhat simplify its properties
slightly. For the Eliashberg function we use a simple square law for $\Omega$%
$<$%
$\Omega_{D}$ and express the prefactor in terms of the electron-phonon
parameter $\lambda$.
\[
\alpha^{2}F\left(  \Omega\right)  =\lambda\left(  \dfrac{\Omega}{\Omega_{D}%
}\right)  ^{2}%
\]
We use for the Debye temperature $\Theta_{D}=80K$ \ and for the Coulomb
parameter $\mu^{\ast}=0.1$. Next we calculate $\lambda\left(  \Omega
_{l}\right)  $ for $\Omega_{l}=l\ast2\pi k_{B}T/\hbar=\omega_{j+l}-\omega_{j}$%
\begin{align*}
\lambda\left(  \Omega_{l}\right)   &  =2\int_{0}^{\Omega_{D}}\lambda\left(
\dfrac{\Omega}{\Omega_{D}}\right)  ^{2}\dfrac{\Omega}{\Omega^{2}+\Omega
_{l}^{2}}d\Omega\\
&  =\lambda\left(  1+\dfrac{\Omega_{l}^{2}}{\Omega_{D}^{2}}\ln\dfrac
{\Omega_{l}^{2}}{\left(  \Omega_{l}^{2}+\Omega_{D}^{2}\right)  }\right)
\end{align*}
These parameters are inserted into equation (\ref{scge}) in the appendix and
yield the dependence of $T_{c}$ on $\lambda$. In Fig.10 this dependence is
plotted, $T_{c}$ as a function of $\lambda$ for $\mu^{\ast}=0.1,$ $\Theta
_{D}=80K$ and $\alpha^{2}F\left(  \Omega\right)  =\lambda\left(  \Omega
/\Omega_{D}\right)  ^{2}$. The transition temperature of Pb, $T_{c}=7.2K$,
corresponds to value of $\lambda_{Pb}=0.75$.
\[%
{\parbox[b]{3.7584in}{\begin{center}
\includegraphics[
height=3.1216in,
width=3.7584in
]%
{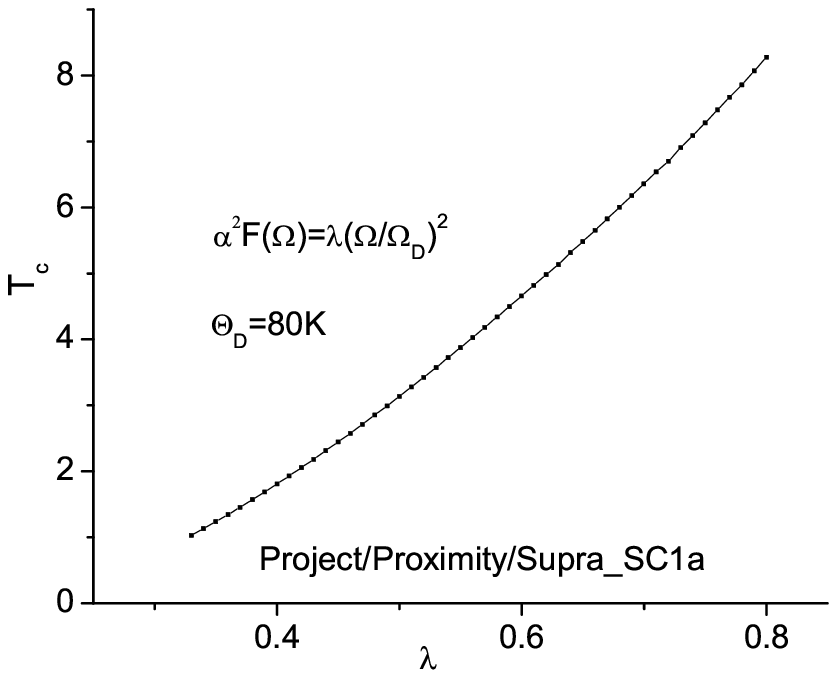}%
\\
Fig.10: The dependence of $T_c $ on the electron-phonon parameter $\lambda$
using the (linear) strong coupling gap equation.
\end{center}}}%
\]

From Fig.10 one obtains $dT_{c}/d\lambda$. At $T_{s}=7.2K$ this slope has the
value $dT_{c}/d\lambda=18.8K$. Within the thin film approximation for
$\lambda$ one finds for the initial slope of $T_{c}$ in a thin double layer:%

\begin{align*}
\dfrac{dT_{c}}{dd_{n}} &  =\dfrac{dT_{c}}{d\lambda}\dfrac{d\lambda}{dd_{n}%
}=18.8\left(  ^{o}K\right)  \dfrac{d}{dd_{n}}\left(  \dfrac{d_{s}N_{s}%
\lambda_{s}+d_{n}N_{n}\lambda_{n}}{d_{s}N_{s}+d_{n}N_{n}}\right) \\
&  =18.8\left(  ^{o}K\right)  \left(  \lambda_{n}-\lambda_{s}\right)
\dfrac{N_{n}}{N_{s}d_{s}}\allowbreak
\end{align*}
or
\[
\dfrac{d_{s}}{T_{s}}\left\vert \dfrac{dT_{c}}{dd_{n}}\right\vert
=2.\,\allowbreak6\,\left(  \lambda_{s}-\lambda_{n}\right)  \dfrac{N_{n}}%
{N_{s}}%
\]

If we take $\lambda_{n}=0.1$ and use $\lambda_{s}=0.75$ then we find
\[
\dfrac{d_{s}}{T_{s}}\left\vert \dfrac{dT_{c}}{dd_{n}}\right\vert
\thickapprox1.7\dfrac{N_{n}}{N_{s}}%
\]

The strong coupling treatment (in the thin film limit) reduces the prefactor
in the intial slope to 1.7. This is a very satisfactory result. The strong
coupling theory yields the correct intial slope. On the other hand it is also
disappointing. It means that the superconducting proximity effect with
superconductors such as Pb one can not be solved with the weak coupling
theory. Furthermore it means that the simulation developed here (as all other
theoretical treatments the author is aware of) can only be applied to weak
coupling superconductors such as Al. But I beleave that this disappointment is
more than compensated by the knowledge that the use of strong coupling
superconductors gives an additional window to the study of strong coupling effects.

\newpage

\section{Conclusion}

This paper derives the transition temperature of a double or multi layer of a
superconductor and a normal conductor numerically. The equivalence in the
propagation of the superconducting pair amplitude and a single electron in
Gorkov's linear gap equation is used. The single electrons act as messengers
who carry the information about the superconducting gap ($N_{s}\Delta\left(
\mathbf{r}^{\prime}\right)  /\tau_{T}$) from one position-time $\left(
\mathbf{r}^{\prime},t^{\prime}<0\right)  $ to another position-time $\left(
\mathbf{r},t=0\right)  $. This message which decays thermally with time as
$\eta_{T}\left(  t\right)  =\sum_{\left\vert \omega_{n}\right\vert <\Omega
_{D}}\exp\left(  -2\left\vert \omega_{n}\right\vert \left\vert t^{\prime
}\right\vert \right)  $, is integrated at $\left(  \mathbf{r,}t=0\right)  $
over all start position-times $\left(  \mathbf{r}^{\prime},t^{\prime}\right)
$ and, after multiplication with the BCS interaction $V_{s},$ \ yields the new
gap function $\Delta\left(  \mathbf{r}\right)  $. At the transition
temperature the procedure has to be self-consistent, i.e. the initial and
final gap function have to be identical. The propagation of the single
electrons is then quasi-classically simulated. The frame work of the
calculation is the weak coupling theory of superconductivity.

This numerical procedure to calculate the transition temperature of double or
multi-layers consisting of thin films of superconductors and normal conductors
is very flexible. The following parameters can be taken from the experiment or
fitted during the calculation:

\begin{itemize}
\item mean free path of the different metals

\item transmission through the interface

\item ratio of specular reflection to diffusive scattering at the surfaces

\item fraction of diffusive scattering at the interface.
\end{itemize}

Furthermore it is possible

\begin{itemize}
\item to vary the mean free path along the thickness of the films

\item to vary the BCS interaction $NV$ at the interface.
\end{itemize}

The few examples which were presented in chapter III demonstrate why the dirty
case is so much simpler than the clean one. They also show that even for small
thickness of the normal metal the gap parameter in the superconductor is not
quite constant. Still the initial slope for an SN double layer follows the
prediction of the Cooper limit.

An important outcome of the numerical simulation is result that the normalized
initial slope of an SN double layer as a function of $d_{n}$ at $d_{n}=0$ does
not depend on

\begin{itemize}
\item the mean free path of the two metal

\item the thickness of the superconductor

\item a (not to large) barrier between the two metal.
\end{itemize}

This slope is essentially given by
\[
\frac{d_{s}}{T_{c0}}\left\vert \frac{dT_{c}}{dd_{n}}\right\vert =\Gamma
_{sn}\frac{N_{n}}{N_{s}}%
\]

For the extreme weak coupling superconductor the value of $\Gamma_{sn}$ is
$1/\left(  NV\right)  _{s}$. If one applies the numeric procedure to double
layers with Pb as the superconducting component then one obtains $\Gamma
_{sn}\thickapprox4.6$. This is in strong disagreement with the results of the
few experiments which allow the evaluation of the initial slope. Their values
for $\Gamma_{sn}$ lie in the range of $1.5-2.0$. The author believes that the
discrepancy is due to the strong coupling properties of the Pb. An analysis of
the strong coupling gap equation in the thin-film limit confirms this
supposition. It yields the value of about $\Gamma_{sn}=1.6$ for a (simplified)
Pb film in contact with a normal metal which is characterized by $\lambda=0.1$
and $\mu^{\ast}=0.1$.

\newpage

\section{Appendix}

\subsection{The kernel in the clean limit}

In the clean limit the thermal Green function has the form
\[
G_{\omega}\left(  \mathbf{r,r}^{\prime}\right)  =-\dfrac{m}{2\pi\hbar
^{2}\left\vert \mathbf{r}-\mathbf{r}^{\prime}\right\vert }\exp\left(
ik_{F}\left\vert \mathbf{r}-\mathbf{r}^{\prime}\right\vert \dfrac{\omega
}{\left\vert \omega\right\vert }-\dfrac{\left\vert \omega\right\vert }{v_{F}%
}\left\vert \mathbf{r}-\mathbf{r}^{\prime}\right\vert \right)
\]
That yields%
\[
H_{\omega}\left(  \mathbf{r,r}^{\prime}\right)  =k_{B}TG_{\omega}\left(
\mathbf{r,r}^{\prime}\right)  G_{\omega}^{\ast}\left(  \mathbf{r,r}^{\prime
}\right)
\]
or%
\[
H_{\omega}\left(  R\right)  =\dfrac{2\pi k_{B}T}{\hbar v_{F}}N\frac{1}{4\pi
R^{2}}\exp\left(  -\dfrac{2\left\vert \omega\right\vert }{v_{F}}R\right)
\]
since $H_{\omega}\left(  \mathbf{r,r}^{\prime}\right)  $ depends only
$R=\left\vert \mathbf{r-r}^{\prime}\right\vert $ (using the BCS-density of
state $N=m^{2}v_{F}/\left(  2\pi^{2}\hbar^{3}\right)  ).$

Without the damping the number of electrons between the radius $R$ and $R+dR$
is
\[
\dfrac{2\pi k_{B}T}{\hbar v_{F}}NdR=\dfrac{2\pi k_{B}T}{\hbar}Ndt
\]
using $dR=v_{F}dt^{\prime}.$ This means that $H_{\omega}\left(  \mathbf{r,r}%
^{\prime}\right)  \Delta\left(  \mathbf{r}^{\prime}\right)  d^{3}%
\mathbf{r}^{\prime}$ corresponds to an injection of
\[
dZ=\dfrac{2\pi k_{B}T}{\hbar}N\Delta\left(  \mathbf{r}^{\prime}\right)
d^{3}\mathbf{r}^{\prime}dt
\]
electrons in the volume $d^{3}\mathbf{r}^{\prime}$ during the time
$dt^{\prime}$ at the position $\mathbf{r}^{\prime}$. $dZ$ is indeed a
(dimensionless) number. The exponential decay $\exp\left(  -2\left\vert
\omega\right\vert R/v_{F}\right)  $ corresponds to a decay with time since
$R=v_{F}t$:
\[
\exp\left(  -2\left\vert \omega\right\vert R/v_{F}\right)  =\exp\left(
-2\left\vert \omega\right\vert t\right)
\]
The density of an electron at the position $\mathbf{r}$ and the time $t=0$
that was injected at $\left(  \mathbf{r}^{\prime},t^{\prime}<0\right)  $ and
propagates with Fermi velocity $v_{F}$ can described by the propagation
density $\rho\left(  v_{F};\mathbf{r,}0;\mathbf{r}^{\prime},t^{\prime}\right)
$. Therefore $H_{\omega}\left(  \mathbf{r},\mathbf{r}^{\prime}\right)  $ can
be written as
\[
H_{\omega}\left(  \mathbf{r},\mathbf{r}^{\prime}\right)  =\dfrac{2\pi k_{B}%
T}{\hbar}N\left(  \mathbf{r}^{\prime}\right)  \int_{-\infty}^{0}dt^{\prime
}\rho\left(  v_{F};\mathbf{r,}0;\mathbf{r}^{\prime},t^{\prime}\right)
\exp\left(  -2\left\vert \omega\right\vert \left\vert t^{\prime}\right\vert
\right)
\]
This yields the gap equation using $\eta_{T}\left(  t^{\prime}\right)  $ from
equation (\ref{dcayf})%
\[
\Delta\left(  \mathbf{r}\right)  =V\left(  \mathbf{r}\right)  \int
d^{3}\mathbf{r}^{\prime}N\left(  \mathbf{r}^{\prime}\right)  \int_{-\infty
}^{0}\dfrac{dt^{\prime}}{\tau_{T}}\rho\left(  v_{F};\mathbf{r,}0;\mathbf{r}%
^{\prime},t^{\prime}\right)  \eta_{T}\left(  t^{\prime}\right)  \Delta\left(
\mathbf{r}^{\prime}\right)
\]

This result applies is not restricted to the clean case but applies to
arbitrary mean free path.\newpage

\subsection{The numerical procedure}

As shown in Fig.2 the metal films are divided in sheets of thickness
$\lambda_{\nu}$. Furthermore the time developement is performed in diffusion
steps of $\tau_{d}=2\lambda_{s}/v_{Fs},$ $t^{\prime}=m\tau_{d}$. Then the
self-consistent gap equation takes the form%

\[
\Delta\left(  z_{\nu}\right)  =\dfrac{V\left(  z_{\nu}\right)  }{\lambda_{\nu
}}\sum_{\nu^{\prime}}\lambda_{\nu^{\prime}}N\left(  z_{\nu^{\prime}}\right)
\sum_{m=0}^{\infty}\dfrac{\tau_{d}}{\tau_{T}}\eta_{T}\left(  m\tau_{d}\right)
\overline{\rho}\left(  z_{\nu}\mathbf{,}m\tau_{d};z_{\nu^{\prime}},0\right)
\Delta\left(  z_{\nu^{\prime}}\right)
\]
(For the zero term in the time summation only half the value is taken). In the
following we denote $\Delta\left(  z_{\nu}\right)  ,V\left(  z_{\nu}\right)
,N\left(  z_{\nu}\right)  $ as $\Delta_{\nu},V_{\nu},N_{\nu}$.

\subsubsection{The value of the BCS interaction $\left(  NV\right)  _{s}$}

For the superconductor with the transition temperature $T_{s}$, the density of
states $N_{s}$ and the Debye temperature $\Theta_{s}$ the implicite equation%

\[
\frac{1}{\left(  NV\right)  _{s}}=\int_{0}^{\infty}\dfrac{dt}{\tau_{T}}%
\eta_{T_{s}}\left(  t/\tau_{T_{s}}\right)
\]
is used.

\subsubsection{Initial conditions}

At the time $t=0$ a simple gap function $\Delta_{\nu}$ in the superconducting
film(s) is chosen, for example $\Delta_{\nu^{\prime}}=k_{B}T_{s}$ for the
superconducting film(s). At the time $t=0$ or $m=0$ we define an occupation
$O_{\nu^{\prime}}\left(  m=0\right)  $ of the different cells%
\[
O_{\nu}\left(  0\right)  =\Delta_{\nu}\lambda_{\nu}N_{\nu}%
\]
This occupation is equally divided in left and right moving electrons
$\overleftarrow{O_{\nu}}\left(  0\right)  $ and $\overrightarrow{O_{\nu}%
}\left(  0\right)  $ with $\overleftarrow{O_{\nu}}\left(  0\right)
=\overrightarrow{O_{\nu}}\left(  0\right)  =O_{\nu}\left(  0\right)  /2$. In
the following sub-sections the recipe is given how to calculate from the
occupation $\overleftarrow{O_{\nu}}\left(  m\right)  ,$ $\overrightarrow
{O_{\nu}}\left(  m\right)  $ at the time $t=m\tau_{d}$ the occupation
$\overleftarrow{O_{\nu}}\left(  m+1\right)  ,$ $\overrightarrow{O_{\nu}%
}\left(  m+1\right)  $. The total occupation is $O_{\nu}\left(  m\right)
=\overleftarrow{O_{\nu}}\left(  m\right)  +\overrightarrow{O_{\nu}}\left(
m\right)  $. With this time devoloping occupation the new gap function becomes%
\[
\widetilde{\Delta_{\nu}}=\dfrac{V_{\nu}}{\lambda_{\nu}}\dfrac{\tau_{d}}%
{\tau_{T}}\sum_{m=0}^{\infty}\eta_{T}\left(  m\tau_{d}\right)  O_{\nu}\left(
m\right)
\]
This iterated gap function has two defects: (i) its shape generally does not
agree with the original gap function $\Delta_{\nu}$, and (ii) the ratio of the
average amplitudes $r=$ $\left\langle \widetilde{\Delta_{\nu}}\right\rangle
/\left\langle \Delta_{\nu}\right\rangle $ will not be one. By determining
numerically $dr/dT$ from two iterations with the same initial gap function and
two temperatures $T$ and $T+T_{\Delta}$ the temperature is adjusted, using
Newton's extrapolation method. After a few iterations $\left\langle
\widetilde{\Delta_{\nu}}\right\rangle $ becomes sufficiently close to
$\left\langle \Delta_{\nu}\right\rangle $ and the adjusted temperature is the
transition temperature of the multi-layer. The iteration is completed when
\[
\dfrac{\sqrt{\dfrac{1}{Z_{s}}\sum_{\nu}\left(  \widetilde{\Delta_{\nu}}%
-\Delta_{\nu}\right)  ^{2}}}{\dfrac{1}{Z_{s}}\sum_{\nu}\Delta_{\nu}}<10^{-5}%
\]
\newpage

\subsection{Diffusive and ballistic propagation}

The important task is to devise a simple fast procedure that describes the
ballistic propagation of the electrons for distances shorter than the mean
free path $l$ and the diffusive propagation for distances larger than $l$. It
helps considerably that only the propagation in z direction has to be modeled
properly (as long as no magnetic field perpendicular to the film is applied).
We consider the electrons in a thin layer of thickness $dz$ in the interval
$\left(  z,z+dz\right)  $. Half of the electrons have a positive z component
$v_{z}=v_{F}\cos\theta$ of the velocity. As long as they are not scattered
their average velocity in z direction is
\[
\left\langle v_{z}\right\rangle =\frac{\int_{0}^{\pi/2}2\pi\sin\theta
v_{F}\cos\theta d\theta}{\int_{0}^{\pi/2}2\pi\sin\theta d\theta}=\frac{1}%
{2}v_{F}%
\]
We take this as the minimum requirement for the ballistic simulation.

The simulation of the diffusion in z direction is rather straight forward. At
the time $t=0$ we have the initial occupation $O_{\nu}\left(  0\right)  $.

Let us first consider the diffusion in one dimension. Here the electrons have
either the velocity $+v_{F}$ or $-v_{F}$. The size of the cells is $\lambda$
and an electron needs the time $\varepsilon_{0}$=$\lambda/v_{F}$ to cross a
cell. We divide the initial occupation $O_{\nu}\left(  0\right)  $ into
$\overleftarrow{O_{\nu}}\left(  0\right)  =$ $\overrightarrow{O_{\nu}}\left(
0\right)  =O_{\nu}\left(  0\right)  /2\ $for the left and right moving
electrons. When the electrons reach the boundary of the cell they will be
partially transmitted through the boundary with the probability $p$ and
partially reflected with the probability $\left(  1-p\right)  $. This yields
the rule how of one obtains from the occupations at the time $t=m\varepsilon
_{0}$ the occupation at the next time step $t=\left(  m+1\right)
\varepsilon_{0}$%
\begin{align*}
\overrightarrow{O_{\nu}}\left(  m+1\right)   & =p\overrightarrow{O_{\nu-1}%
}\left(  m\right)  +\left(  1-p\right)  \overleftarrow{O_{\nu}}\left(
m\right) \\
\overleftarrow{O_{\nu}}\left(  m+1\right)   & =p\overleftarrow{O_{\nu+1}%
}\left(  m\right)  +\left(  1-p\right)  \overrightarrow{O_{\nu}}\left(
m\right)
\end{align*}

This yields a one-dimensional diffusion with the diffusion constant
$D=\frac{1}{2}\frac{p}{1-p}\frac{\lambda^{2}}{\varepsilon_{0}}$.

Ballistic propagation requires setting $p$ almost equal to 1. In this case
almost all the $\overrightarrow{O_{\nu}}\left(  m\right)  $ electrons move
from cell $\nu$ to cell $\left(  \nu+1\right)  $ during the time
$\varepsilon_{0}$. This means that they propagate the average distance
$\lambda=v_{F}\varepsilon_{0}$ during the time $\varepsilon_{0}$. Therefore
this model does not fulfill the basic requirement for ballistic propagation in
three dimension that $\ \left\langle v_{z}\right\rangle =\frac{1}{2}v_{F}$.

A three-dimensional diffusion can be obtained by a sequential propagation in
x, y and z direction, each for a time of $\varepsilon_{0}$ with the velocity
$v_{F}$. This yields a diffusion constant $D=\frac{1}{2}\frac{p}{1-p}%
\frac{\lambda^{2}}{3\varepsilon_{0}}$ and triples the average time for the
diffusion in z direction. Since the electrons propagate only during every
third of the interval $3\varepsilon_{0}$ in z direction they propagate the
distance\ $\lambda$ during the time $3\varepsilon_{0}$, i.e. their average
velocity in z direction is only $\left\langle v_{z}\right\rangle =v_{F}/3$.

We can simulate the average diffusive and ballistic propagation of the
electrons in z direction by propagating every other time interval
$\varepsilon_{0}$ in z direction. Then the time step is $\tau_{d}%
=2\varepsilon_{0}$. In this case the diffusion constant is $D=\frac{1}{2}%
\frac{p}{1-p}\frac{\lambda^{2}}{\tau_{d}}$ and the ballistic propagation
yields $\left\langle v_{z}\right\rangle =v_{F}/2$ as required.

It should be mentioned that it is essential that the electron density is
divided into (at least) two components, one for motion in the $+z$ and the
other for -z \ direction. A single density component with hopping to neighbor
places yields only small diffusion constants of $D=\frac{p}{2}\frac
{\lambda^{2}}{\varepsilon_{0}}$ and can't describe the ballistic propagation
at all.

For the normal conductor the same time element $\tau_{d}$ is used to simulate
the propagation. The thickness $d_{n}$ is divided in cells (or layers) of
thickness $\lambda_{n}=v_{F,n}\varepsilon_{0}=v_{F,n}\tau_{d}/2 $. This
synchornizes the diffusion in the whole double layer.

The transparency $p$ of the cell walls is obtained form the experimental
conductivity $\sigma$ of the films, where $\sigma_{m}=2e^{2}N_{m}D_{m}$ or%
\[
D_{m}=\dfrac{\sigma_{m}}{2e^{2}N_{m}}\text{, }p_{m}=\dfrac{D_{m}}{\left(
\frac{1}{2}\frac{\lambda_{m}^{2}}{2\varepsilon_{0}}+D_{m}\right)  }%
\]
where $m$ stands for $s$ or $n$.

\newpage

\subsection{Interface between two films}

The transmission of electrons through an interface between two metals (which
we denote with S and N) is only in exceptional cases equal to 1. If for
example the Fermi wave number $k_{F,s}$ is larger then $k_{F,n}$ then any
electron in S whose component $k_{\rho}$ parallel to the surface is larger
than $k_{F,n}$ cannot cross the interface because afterwards it would have an
energy of a least $\left(  \hbar k_{\rho}\right)  ^{2}/2m$ which is larger
than the Fermi energy $E_{F,n}=\left(  \hbar k_{F_{,n}}\right)  ^{2}/2m$ in
the normal conductor. An electron in N with Fermi energy would not violate the
conservation of energy when crossing the interface. However, a plane wave
which crosses a step in the potential energy is partially reflected. Therefore
the transition probability is less than 1 for any electron. If one averages
the transition probability of all these electrons (to cross the interface from
N to S) one finds
\begin{align*}
T_{N->S}  & =f\left(  \frac{E_{F,n}}{E_{F,s}}-1\right)  \text{, where}\\
f\left(  x\right)   & =\frac{4}{15}\frac{\left(  \sqrt{\left(  x+1\right)
}\right)  ^{3}\left(  x+6\right)  -\left(  \sqrt{x}\right)  ^{5}-10x-6}{x^{2}}%
\end{align*}
For small $x$ the asymptotic expansion is $f\left(  x\right)  \backsimeq
\left(  1-\frac{4}{15}\sqrt{x}\right)  $.

The detailed balance requires that in equilibrium the number of electrons
which cross from S to N is equal to the number of electron which cross from N
to S. Let us assume that the electron distribution is in equilibrium and we
consider an interface S/N. $O_{s}\left(  m\right)  $ and $O_{n}\left(
m\right)  $ are the occupations in the cells on the left and right side of the
interface. The transmission coefficients are by $T_{sn}$ and $T_{ns}$. Then
the occupation at the time $\left(  m+1\right)  \tau_{d}$ is%
\begin{align*}
\overleftarrow{O_{s}}\left(  m+1\right)   & =T_{ns}\overleftarrow{O_{n}%
}\left(  m\right)  +\left(  1-T_{sn}\right)  \overrightarrow{O_{s}}\left(
m\right) \\
\overrightarrow{O_{n}}\left(  m+1\right)   & =T_{sn}\overrightarrow{O_{s}%
}\left(  m\right)  +\left(  1-T_{ns}\right)  \overleftarrow{O_{n}}\left(
m\right)
\end{align*}
In equilibrium one has $\overleftarrow{O_{s,n}}=\overrightarrow{O_{s,n}%
}=\dfrac{1}{2}O_{s,n}$ and $O_{s,n}\left(  m+1\right)  =O_{s,n}\left(
m\right)  $. This yields
\[
T_{ns}O_{n}\left(  m\right)  =T_{sn}O_{s}\left(  m\right)
\]
Since $O_{s,n}=\lambda_{s,n}N_{s,n}$ one obtains finally
\[
\dfrac{T_{sn}}{T_{ns}}=\dfrac{\lambda_{n}N_{n}}{\lambda_{s}N_{s}}%
\]

If one considers real metals a considerably more difficult situation arises
when the superconductor has a mass enhancement of the density of states (as
most superconductor have, in particular the strong coupling ones). We return
to the mass enhancement below. However, independent how complicated the
individual transmission probabilities are, the detailed balance will always
apply. In our simulation we use $T_{ns}\leq1$ as a fit parameter and calculate
$T_{sn}$ using the detailed balance.\newpage

\subsection{Strong coupling effects}

In appendix (6.1) the kernel $H_{\omega}\left(  \mathbf{r,r}^{\prime}\right)
$was derived for free electrons
\[
H_{\omega}\left(  \mathbf{r,r}^{\prime}\right)  =k_{B}TG_{\omega}\left(
\mathbf{r,r}^{\prime}\right)  G_{\omega}^{\ast}\left(  \mathbf{r,r}^{\prime
}\right)
\]
where
\[
G_{\omega}\left(  \mathbf{r,r}^{\prime}\right)  =-\dfrac{m}{2\pi\hbar
^{2}\left\vert \mathbf{r}-\mathbf{r}^{\prime}\right\vert }\exp\left(
ik_{F}\left\vert \mathbf{r}-\mathbf{r}^{\prime}\right\vert \dfrac{\omega
}{\left\vert \omega\right\vert }-\dfrac{\left\vert \omega\right\vert }{v_{F}%
}\left\vert \mathbf{r}-\mathbf{r}^{\prime}\right\vert \right)
\]
$G_{\omega}\left(  \mathbf{r,r}^{\prime}\right)  $ is a Fourier transform of
$G_{\omega}\left(  \mathbf{k}\right)  $
\[
G_{\omega}\left(  \mathbf{k}\right)  =\dfrac{1}{i\omega-\varepsilon_{k}}%
\]
In the renormalized strong coupling case one has%
\[
G_{\omega}\left(  \mathbf{k}\right)  =\dfrac{1}{1+\lambda}\dfrac{1}%
{i\omega-\varepsilon_{k}}%
\]
where $\left(  1+\lambda\right)  $ is the "mass enhancement" of the electrons
at the Fermi surface due to the electron-phonon interaction. Here $\lambda$ is
defined as%
\[
\lambda=2\int_{0}^{\Omega_{D}}\alpha^{2}F\left(  \omega\right)  \dfrac
{d\omega}{\omega}%
\]
and $\alpha^{2}F\left(  \omega\right)  $ is the Eliashberg function of the
electron-phonon interaction. In performing the Fourier transform one obtains%
\[
G_{\omega}\left(  R\right)  =\dfrac{m}{2\pi\hbar^{2}}\dfrac{1}{R}\exp\left[
ik_{F}R\dfrac{\omega}{\left\vert \omega\right\vert }-\dfrac{\left\vert
\omega\right\vert }{\hbar v_{F}^{\ast}}R\right]
\]
(using $\left\vert \mathbf{r-r}^{\prime}\right\vert =R$). Compared with the
free electron Green function the Fermi velocity is now renormalized.

This yields for the function $H_{\omega}\left(  R\right)  $
\[
H_{\omega}\left(  R\right)  =\dfrac{2\pi k_{B}T}{\hbar}\dfrac{N}{v_{F}}%
\dfrac{1}{4\pi R^{2}}\exp\left(  -\dfrac{2\left\vert \omega\right\vert }{\hbar
v_{F}^{\ast}}R\right)
\]
using the bare BCS-density of state $N=m^{2}v_{F}/2\pi^{2}\hbar^{3}$ and the
bare Fermi velocity $v_{F}$.

Now we use the same argument as before: The term $H_{\omega}\left(
\mathbf{r,r}^{\prime}\right)  \Delta\left(  \mathbf{r}^{\prime}\right)
d^{3}\mathbf{r}^{\prime}$ corresponds to an injection of
\[
dZ=\dfrac{1}{\left(  1+\lambda\right)  }\dfrac{2\pi k_{B}T}{\hbar}%
N\Delta\left(  \mathbf{r}^{\prime}\right)  d^{3}\mathbf{r}^{\prime}dt
\]
electrons in the volume $d^{3}\mathbf{r}^{\prime}$ during the time $dt$ at the
position $\mathbf{r}^{\prime}$ which propagate with $v_{F}^{\ast}$. The factor
$1/\left(  1+\lambda\right)  $ arrives from the ratio of $v_{F}^{\ast}/v_{F}$.
The exponential decay $\exp\left(  -2\left\vert \omega\right\vert
R/v_{F}^{\ast}\right)  $ corresponds to a decay with time as $\exp\left(
-2\left\vert \omega\right\vert R/v_{F}^{\ast}\right)  =\exp\left(
-2\left\vert \omega\right\vert t^{\prime}\right)  $ since $R=v_{F}^{\ast
}t^{\prime}$ . Now we replace the BCS interaction $V$ by an effective
interaction strength $\left(  \lambda-\mu^{\ast}\right)  /N$. \ ($\alpha
^{2}F\left(  \omega\right)  $ and therefore $\lambda$ contains the bare
density of states $N$ as a factor)

The gap equation is then
\[
\Delta\left(  \mathbf{r}\right)  =\dfrac{\left(  \lambda\left(  \mathbf{r}%
\right)  -\mu^{\ast}\right)  }{N\left(  \mathbf{r}\right)  }\frac{2\pi k_{B}%
T}{\hbar}\int d^{3}\mathbf{r}^{\prime}\dfrac{N\left(  \mathbf{r}^{\prime
}\right)  }{\left(  1+\lambda\left(  \mathbf{r}^{\prime}\right)  \right)
}\int_{-\infty}^{0}\rho\left(  v_{F}^{\ast};\mathbf{r,}0;\mathbf{r}^{\prime
},t^{\prime}\right)  \eta_{T}\left(  t^{\prime}\right)  dt^{\prime}%
\Delta\left(  \mathbf{r}^{\prime}\right)
\]
A test for constant $\Delta$ using $\int d^{3}\mathbf{r}^{\prime}\rho\left(
v_{F}^{\ast};\mathbf{r,}0;\mathbf{r}^{\prime},t^{\prime}\right)  =1$ yields%
\[
\Delta=\dfrac{\left(  \lambda-\mu^{\ast}\right)  }{N}\frac{2\pi k_{B}T}{\hbar
}\dfrac{N}{1+\lambda}\int_{-\infty}^{0}\eta_{T_{c}}\left(  t^{\prime}\right)
dt^{\prime}\Delta
\]
This yields McMillan's first approximation for the $T_{c}$ of a strong
coupling superconductor
\begin{align*}
\dfrac{\left(  1+\lambda\right)  }{\left(  \lambda-\mu^{\ast}\right)  }  &
=\sum_{n=0}^{\Omega_{D}\tau_{T_{c}}}\dfrac{1}{n+\dfrac{1}{2}}\\
T_{c}  & \thickapprox\left\langle \Omega\right\rangle \exp\left(
\dfrac{\left(  1+\lambda\right)  }{\left(  \lambda-\mu^{\ast}\right)
}\right)
\end{align*}

\subsubsection{Strong coupling gap equations}

For the strong coupling superconductor one has to use the Eliashberg theory.
It replaces the "one" gap equation by series of gap equations at different
Matsubara frequencies. Equation (\ref{scge}) \cite{B86} defines the gap
$\Delta\left(  \omega_{l}\right)  $ at $\omega_{l}$ as a function of the gap
at all other Matsubara frequencies $\omega_{j}$.
\begin{align}
\Delta\left(  \omega_{i}\right)   & =\dfrac{2\pi k_{B}T}{\hbar}\sum
_{j}\left\{  \lambda\left(  \omega_{i}-\omega_{j}\right)  -\mu^{\ast}\right\}
\dfrac{1}{2\left\vert \widetilde{\omega_{j}}\right\vert }\Delta\left(
\omega_{j}\right) \label{scge}\\
\widetilde{\omega_{j}}  & =\omega_{j}+\dfrac{\pi k_{B}T}{\hbar}\sum_{l}%
\dfrac{\omega_{l}}{\left\vert \omega_{l}\right\vert }\lambda\left(  \omega
_{j}-\omega_{l}\right) \nonumber\\
\lambda\left(  \omega_{i}-\omega_{j}\right)   & =2\int_{0}^{\infty}%
d\omega\dfrac{\omega\alpha^{2}F\left(  \omega\right)  }{\omega^{2}+\left(
\omega_{i}-\omega_{j}\right)  ^{2}}\nonumber
\end{align}

The effective BCS interaction $NV$ is replaced by $\left\{  \lambda\left(
\omega_{i}-\omega_{j}\right)  -\mu^{\ast}\right\}  $. The electron-phonon
parameter $\lambda\left(  \omega_{i}-\omega_{j}\right)  $ is determined by the
Eliashberg function $\alpha^{2}F\left(  \omega\right)  $. The denominator
$1/2\left\vert \omega_{n}\right\vert $ is replaced by the dressed Matsubara
frequencies $1/2\left\vert \widetilde{\omega_{j}}\right\vert $. Here is is
$\widetilde{\omega_{j}}=\omega_{j}\left(  1+\overline{\lambda}\right)  $ where
$\overline{\lambda}$ is an average over $\left(  2j+1\right)  $ different
values of $\lambda\left(  \Omega_{\nu}\right)  $, ($\Omega_{\nu}=\nu2\pi
k_{B}T/\hbar,0\leq\nu\leq j$).

One can derive from the energy dependent gap equation a renormalized gap
equation by the following simplifications

\begin{itemize}
\item $\Delta\left(  \omega_{i}\right)  \Rightarrow\Delta=const$ for
$\left\vert \omega_{i}\right\vert <\Omega_{D}$

\item $\Delta\left(  \omega_{i}\right)  \Rightarrow0$ for $\left\vert
\omega_{i}\right\vert >\Omega_{D}$

\item one replaces $\lambda\left(  \omega_{i}-\omega_{j}\right)
\Rightarrow\lambda\left(  0\right)  =\lambda$ for $\left\vert \omega
_{i}\right\vert ,\left\vert \omega_{j}\right\vert <\Omega_{D}$
\end{itemize}

Then one finds that $\widetilde{\omega_{j}}$ is renormalized as $\widetilde
{\omega_{j}}=\omega_{j}\left(  1+\lambda\right)  $ where $\left(
1+\lambda\right)  $ is the electron-phonon renormalization factor. Using these
simplifications the renormalized gap equation takes the form%
\[
\Delta=\dfrac{2\pi k_{B}T}{\hbar}\sum_{\left\vert \omega_{j}\right\vert
\leq\Omega_{D}}\dfrac{\lambda-\mu^{\ast}}{1+\lambda}\dfrac{1}{2\left\vert
\omega_{j}\right\vert }\Delta
\]
as before.

\end{document}